\documentclass{aastex}
\usepackage{emulateapj5,apjfonts,psfig}

\newcommand{\etal}{{et~al.}}
\newcommand{\kms}{{km~s$^{-1}$}}

\begin{document}
 
\title{The DEEP Groth Strip Survey IX: Evolution of the Fundamental
Plane of Field Galaxies}

\author{Karl Gebhardt\altaffilmark{1,2,3}, S.M. Faber\altaffilmark{2},
David C. Koo\altaffilmark{2}, Myungshin Im\altaffilmark{2}, Luc
Simard\altaffilmark{4}, Garth D. Illingworth\altaffilmark{2}, Andrew
C. Phillips\altaffilmark{2}, Vicki L. Sarajedini\altaffilmark{5},
Nicole P. Vogt\altaffilmark{6}, Benjamin Weiner\altaffilmark{2}, Christopher
N. A. Willmer\altaffilmark{2,7}}

\altaffiltext{1}{Hubble Fellow}

\altaffiltext{2}{University of California Observatories / Lick
Observatory, University of California, Santa Cruz, CA 95064}

\altaffiltext{3}{Present Address: Astronomy Department, University
of Texas, Austin, TX 78723}

\altaffiltext{4}{Herzberg Institute of Astrophysics, National Research
Council of Canda, 5071 West Saanich Road, Vistoria, BC V9E 2E7,
Canada}

\altaffiltext{5}{Department of Astronomy, University of Florida,
Gainesville, FL 32611-2055}

\altaffiltext{6}{Department of Astronomy, New Mexico State
university, Las Cruces, NM 88003-8001}

\altaffiltext{7}{On leave from Observatorio Nacional Brazil}

\begin{abstract}

Fundamental Plane studies provide an excellent means of understanding
the evolutionary history of early-type galaxies. Using the Low
Resolution Imaging Spectrograph on the Keck telescope, we obtained
internal stellar kinematic information for 36 field galaxies in the
Groth Strip---21 early-type and 15 disk galaxies. Their redshifts
range from 0.3--1.0, with a median redshift 0.8. The slope of the
relation shows no difference compared with the local slope. However,
there is significant evolution in the zero-point offset; an offset due
to evolution in magnitude requires a 2.4 magnitude luminosity
brightening at $z$=1. We see little differences of the offset with
bulge fraction, which is a good surrogate for galaxy type. Correcting
for the luminosity evolution reduces the orthogonal scatter in the
Fundamental Plane to 8\%, consistent with the local scatter. This
scatter is measured for our sample, and does not include results from
other studies which may have different selection effects. The
difference in the degree of evolution between our field sample and
published cluster galaxies suggests a more recent formation
epoch---around $z=1.5$ for field galaxies compared to $z>2.0$ for
cluster galaxies. The magnitude difference implies that the field
early-type galaxies are about 2 Gyr younger than the cluster
ellipticals using standard single-burst models. However, the same
models imply a significant change in the rest-frame $U-B$ color from
then to present, which is not seen in our sample. Continuous low-level
star formation, however, would serve to explain the constant colors
over this large magnitude change. A consistent model has 7\% of the
stellar mass created after the initial burst, using an exponentially
decaying star formation rate with an e-folding time of 5 Gyr.

\end{abstract}

\section{Introduction}

As we probe to higher redshifts, changes in the structural properties
of galaxies provide necessary information to understand their
formation and evolution. The Fundamental Plane (FP) is one of the main
correlations that reflect the internal structure of early-type
galaxies, and so has become a primary tool for studying galaxy
evolution. Since the Fundamental Plane relates sizes, magnitudes, and
velocity dispersions, changes in relative positions and the density
distribution within the Fundamental Plane not only allow us to
determine which evolutionary processes are dominant at early times,
but also provide important information as to the formation epoch. All
three properties may have evolutionary signatures: galaxies may grow
in size due to accretion, change their brightness due to the stellar
population, and even alter their projected internal dispersions due to
evolution in the orbital structure or mass growth. The most important
evolutionary effect is likely the fading of galaxies as the stellar
population ages; based on current formation times for stars in
elliptical galaxies---generally accepted to be before $z$=2---we
should then expect substantial fading from $z$=1 to the present. In
fact, many groups measure values consistent with this picture:
van~Dokkum~\etal\ (1998) find almost one magnitude of evolution in the
surface brightness for a cluster at $z$=0.83, while Bender~\etal\
(1998), Kelson~\etal\ (1997), and J\o rgensen~\etal\ (1999) find
consistent numbers for clusters at lower redshifts. The evidence to
date points to a formation epoch for cluster ellipticals at $z>2$.


\begin{figure*}[t]
\centerline{\psfig{file=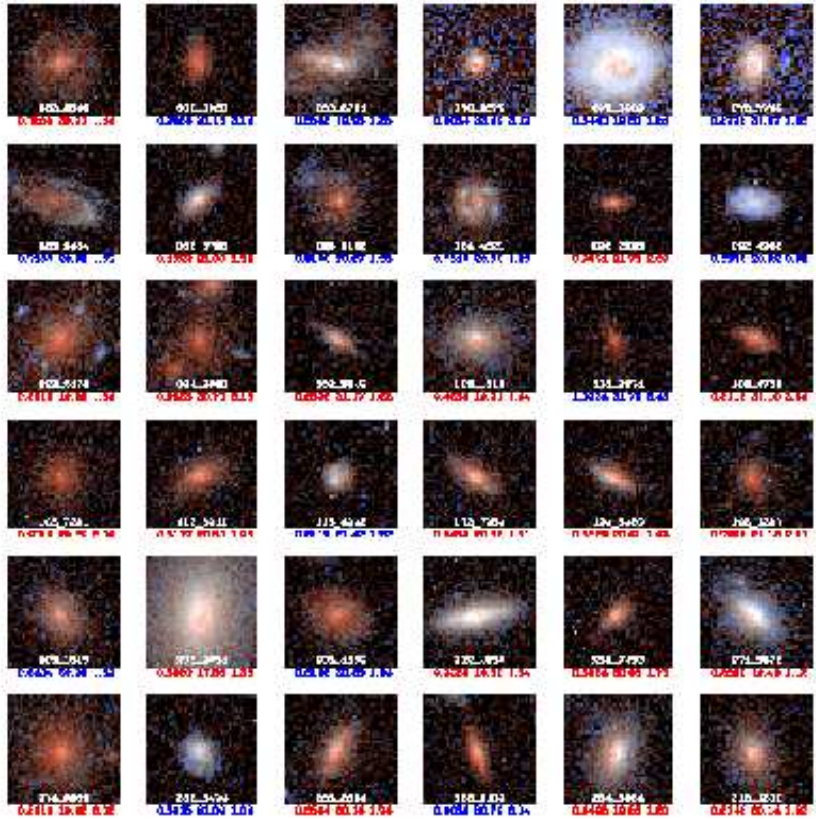,width=14cm,angle=0}}
\figcaption[gebhardt.fig1.ps]{HST real color images of the central
3\arcsec x3\arcsec\ for the 36 galaxies. The numbers below each image
refer to the redshift, $I$ total magnitude, and $V-I$. The colors
represent a logarithmic stretch using the $V$ and $I$ images only. If
the text on the bottom of an image is red, then that galaxy is an
early-type (from Im~\etal\ 2002), otherwise the text is blue.
High-resolution version available at 
http://hoku.as.utexas.edu/$_{^\sim}$gebhardt/FPpaper.ps.
\label{fig1}}
\end{figure*}


Cluster galaxies likely represent one of the earliest populations to
form; because they exist in high-density regions that rapidly
condensed out of the initial mass fluctuations, we expect them to show
the oldest stellar populations. Galaxies in the field require longer
timescales for their formation, and we might therefore expect to see
differences between field and cluster galaxies in their Fundamental
Plane properties. Comparison of the two samples provides a relative
timescale for galaxy formation that depends on environment, which
helps to constrain cosmological models. Our goal is to explore the
properties of a well defined sample of field galaxies. At redshifts
around one, the predicted age difference between field and cluster
galaxies is around 3Gyr (Kauffmann 1996), which, for single-burst
stellar populations, would translate into a one-magnitude difference
in offset from the Fundamental Plane.

There are other groups measuring the properties of field ellipticals.
Van~Dokkum~\etal\ (2001) present 18 galaxies out to a redshift of 0.55
and find slightly more evolution in the field galaxies compared to
their cluster sample, but the small differences only imply a very
modest formation age difference of 1 Gyr or less. Van~Dokkum \& Ellis
(2003) extend this analysis with field galaxies in the North Hubble
Deep Field out to redshift of one, and also find very little
difference between cluster and field galaxies. Treu~\etal\ (2002)
study the FP evolution from 30 field galaxies between redshifts 0.1 to
0.66 and find an evolution in the luminosity larger than what has been
found for the cluster samples, consistent with a much more recent
formation epoch around $z<2$. However, Koopmans \& Treu (2002) provide
one galaxy at redshift 1.0 which has a magnitude offset in the FP
consistent with the cluster galaxy offsets. Thus, there is no clear
picture of the differences, if any, between cluster and field
ellipticals. One problem is that at redshifts less than 0.7, the
expected magnitude differences are small even for galaxies with
formation epochs separated by a few Gyr. The passive luminosity
evolution for a galaxy is greatest when the galaxy is less than 2 Gyrs
old, and the amount of evolution rapidly decreases after that. Thus,
the magnitude difference going from 4 to 7 Gyr is significantly less
than that going from 1 to 4 Gyr. Given the uncertainties and intrinsic
scatter in FP studies, it is imperative to push the data to as high a
redshift as possible to see the differences. This paper presents
results for 36 galaxies between $0.3<z<1.1$, with a median redshift of
0.8.

The DEEP (Deep Extragalactic Evolutionary Probe) team has been
obtaining high signal-to-noise, high resolution data for a
magnitude-limited sample of field galaxies in the Groth Survey Strip,
taken under HST programs GTO 5090 (PI: Groth) and GTO 5109 (PI:
Westphal).  The goal is to study the morphological and kinematic state
of galaxies at redshifts from 0.5 to 1.0. This project is a long-term
endeavor, and initial results have provided an important increase in
our understanding of disk galaxy formation and evolution (Vogt~\etal\
1996; Simard~\etal\ 1999). In this paper, we present {\it
absorption-line} kinematic results for galaxies in the DEEP
sample. Over the next few years, this sample will increase
dramatically. Throughout this paper we use a cosmology of H$_0=70$,
$\Omega_{\rm m}=0.3$ and $\Omega_\Lambda=0.7$.

\section{Observations and Reductions}

The photometric observations consist of HST $V$ and $I$ band images,
with exposure times of 2800s in F606W and 4400s in F814W. The full
field contains 28 overlapping {\it HST} pointings located at
approximately 1417+52, called the Groth Survey Strip. The area spans
50\arcmin\ by 2.5\arcmin. Simard~\etal\ (2002) provide details for the
reduction and photometry. The full dataset consists of 7450 galaxies
with structural parameters determined from a two-dimensional
bulge+disk surface brightness model (Simard~\etal\ 2002). The data in
this paper span the full angular length of the Groth Strip, but the
density of coverage is not complete.

We obtained spectroscopic observations over several years, from 1995
to 2000, on the Keck telescope using the Low Resolution Imaging
Spectrograph, LRIS (Oke~\etal\ 1995). Two gratings were used to obtain
large wavelength coverage: a blue side, which ranges roughly from
4500~\AA\ to 6500~\AA, and a red side, ranging from 6000~\AA\ to
9500~\AA. The exact wavelength range depends on the slit's position in
the mask. For the blue side, the spectral scale is 0.84\AA~pix$^{-1}$,
and for the red side it is 1.2\AA~pix$^{-1}$. The spatial scale for
both is 0.215\arcsec~pix$^{-1}$. Both the slit width and the seeing
are typically 1\arcsec.

The spectroscopic reduction follows a pipeline procedure described in
Weiner~\etal\ (2003). We combine all spectra for a particular galaxy;
in some cases we combined spectra taken as much as five years
apart. Typically, the early-type galaxies appeared on several mask
designs since we desired higher signal-to-noise (S/N) for
absorption-line dispersions. The number of individual exposures ranges
from two to fifteen, with total exposure times ranging from 35 to 310
minutes. Table~1 presents the galaxy sample, along with the exposure
times.

\subsection{Object Selection}

There are three selection steps that determine the final sample.
These are the galaxy identifications from the {\it HST} image, the
spectroscopic selection from that sample, and finally the selection of
galaxies from the spectroscopic sample that go into this analysis. The
photometric selection of the galaxies is described in detail in
Simard~\etal\ (2002). For the analysis in this paper, this selection
effect is negligible since there are no photometric biases down to
$I=23.5$, and our faintest galaxy is $I=22.3$. The original
spectroscopic sample consists of 648 galaxies. It is magnitude-limited
with $I<23$ as galaxy candidates. In order to provide dense spatial
coverage of the GSS fields, we used many different mask designs per
field. We have obtained spectra in 25 of the 28 GSS fields, but in
each field the completeness varies significantly. In some of the
fields, we have spectra for all $I<23$ galaxies, while in others the
fraction observed is below 20\%. 

The selection criteria for the absorption-line studies depend only on
S/N and the presence of absorption-lines; we measure internal
absorption-line kinematics for all objects with adequate
signal-to-noise ($>5$ per pixel), regardless of galaxy type. Since we
require absorption-lines, the sample consists of predominantly
early-type galaxies. Im~\etal\ (2002) identify early-type galaxies
from the DEEP sample based on structural parameters; 21 galaxies
(58\%) from our sample are classified as early type galaxies in that
paper. Im~\etal\ (2002) provide a detailed description of the
separation into early and late type galaxies. This separation relies
primarily on bulge fraction ($>$ 0.4), smoothness and symmetry, but
not color. We refer the reader to that work for more information on
the classification system. The remaining 15 galaxies in our sample
(42\%) are likely disk galaxies, but we measure them anyway; their
results will be discussed separately. Im~\etal\ have an additional 22
galaxies which we do not include since they are either too faint or
have absorption lines that fall into unusable spectral regions due to
strong night sky emission. Koo~\etal\ (2003) study bulge properties of
the DEEP sample and include all of our galaxies.  Figure~1 presents
true color images for the whole sample. In the following, we make a
clear distinction between early-types and disk-dominated systems, and
we will discuss each separately.

Of the three selection effects---the original photometric sample
selection, spectroscopic selection, and absorption line selection
effect---the ability to measure absorption lines causes the most
severe constraint. The photometric catalogue contains every galaxy
observed with {\it HST} in the F814W and F606W filters down to
$I_{814}$=24 and $V_{606}$=25. Since our faintest galaxy in the FP
sample is $I=22.3$, the photometric selection causes no bias (for
further description of the photometric selection function, see
Simard~\etal\ 2002). Since our faintest galaxy is relatively bright
compared to the total spectroscopic sample, the spectroscopic
selection function also has little effect on the overall selection
function. Koo~\etal\ (2003) discuss the selection function for bulges
(which include the galaxies in this sample) and find that the most
important effect is the magnitude of the bulge; i.e., size,
bulge-to-total ratio, and color have small effect on the spectroscopic
selection function at our magnitude limit of $I=22.3$. Thus, by far,
the most important selection effect is the ability to measure
dispersions from the absorption lines. We find that this selection
does not bias our results, as discussed in \S3.


\begin{figure*}[t]
\vskip -30pt
\centerline{\psfig{file=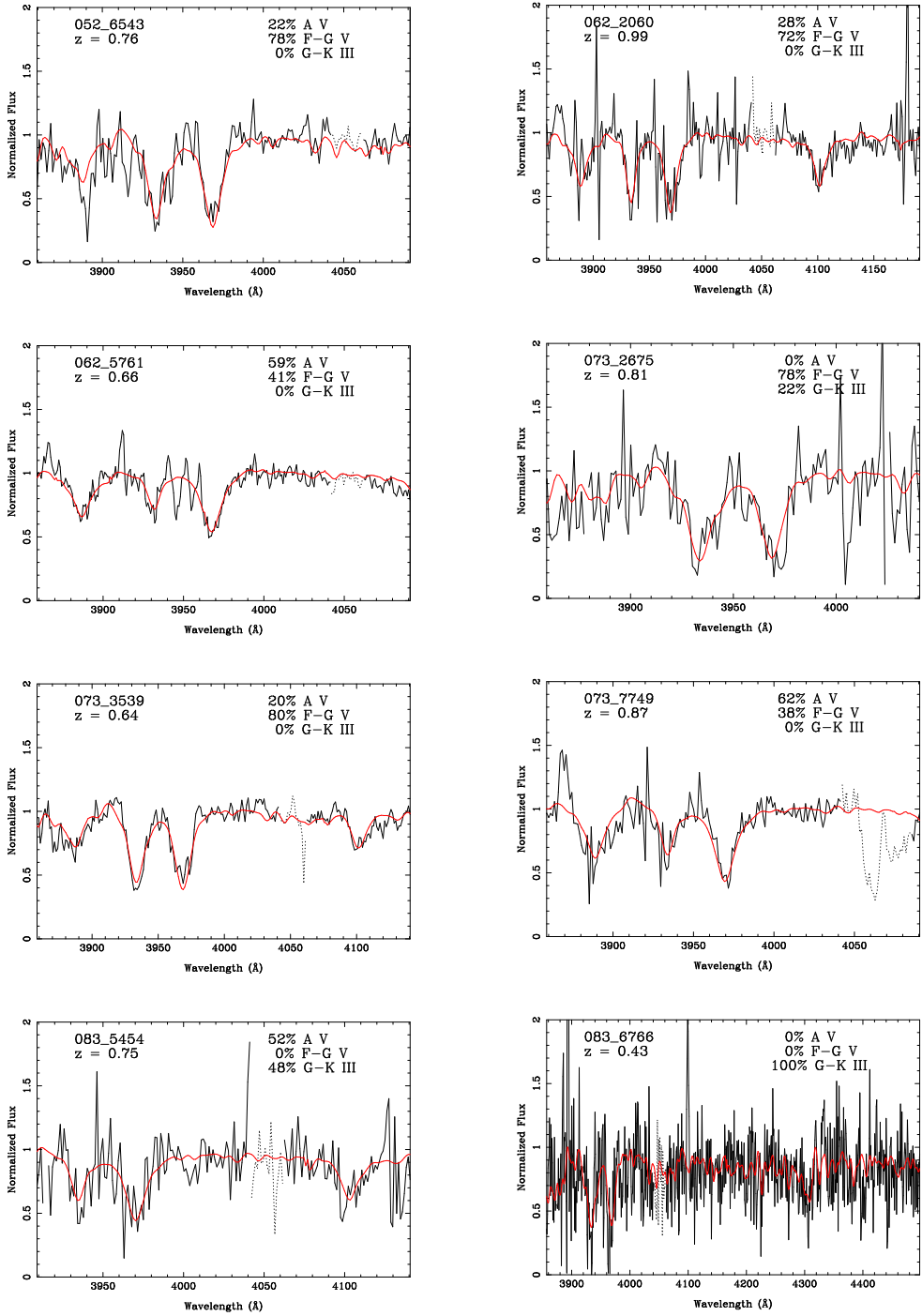,width=16cm,angle=0}}
\vskip -50pt \figcaption[sp1.ps]{Spectra for the sample of
galaxies. In each panel, the noisy black line represents the spectral
data; each galaxy has considerably more spectral coverage but we show
only the region used for the kinematic analysis. Dotted lines in the
spectra indicate those regions that were excluded from the fitting due
to night sky contamination. The red line is the velocity profile
convolved with the best-fit weighted averaged template, as shown in
the upper-right corner.
\label{fig2}}
\end{figure*}

\begin{figure*}[t]
\figurenum{2}
\vskip -30pt
\centerline{\psfig{file=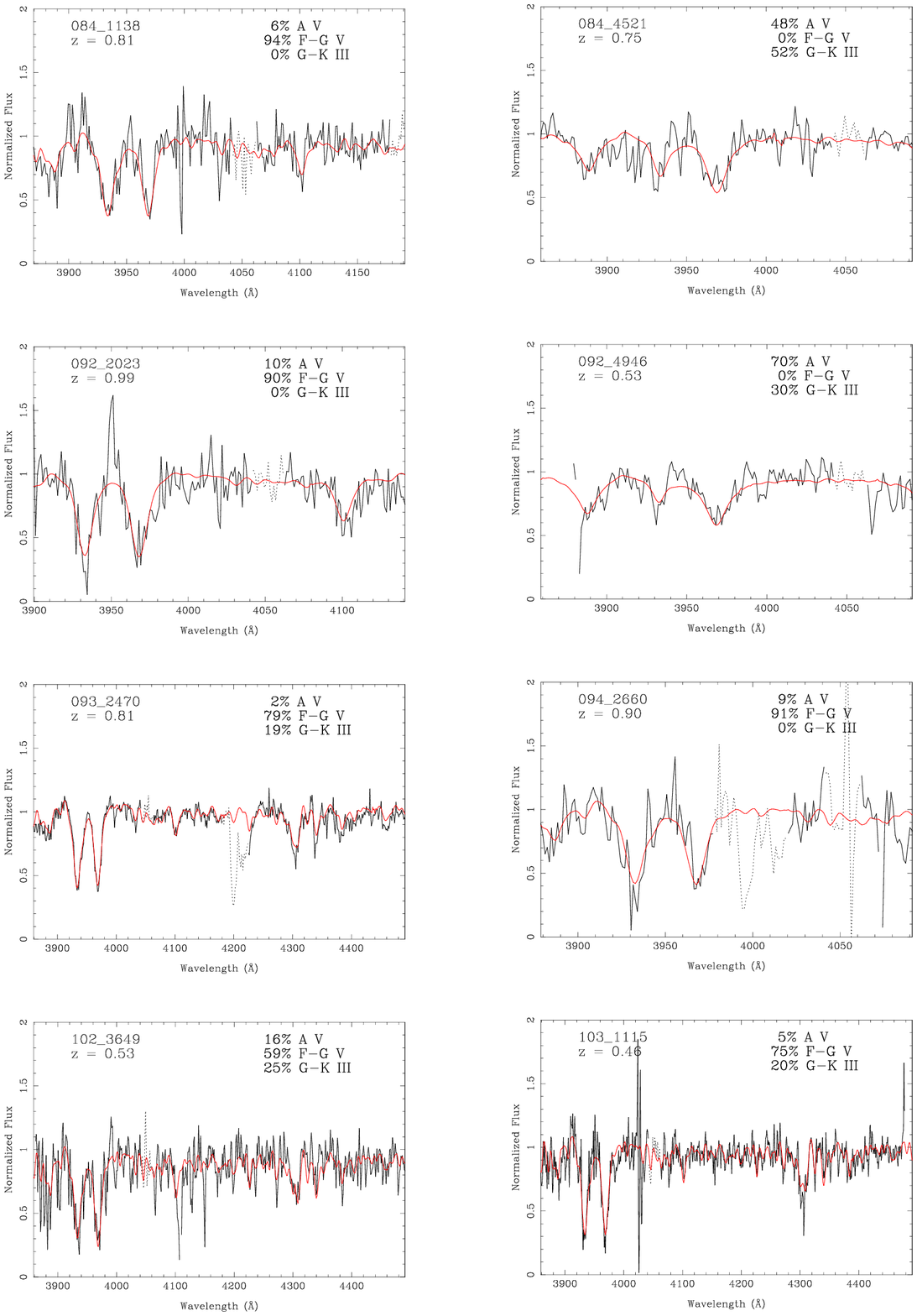,width=16cm,angle=0}}
\vskip -50pt
\figcaption{continued}
\end{figure*}

\begin{figure*}[t]
\figurenum{2}
\vskip -30pt
\centerline{\psfig{file=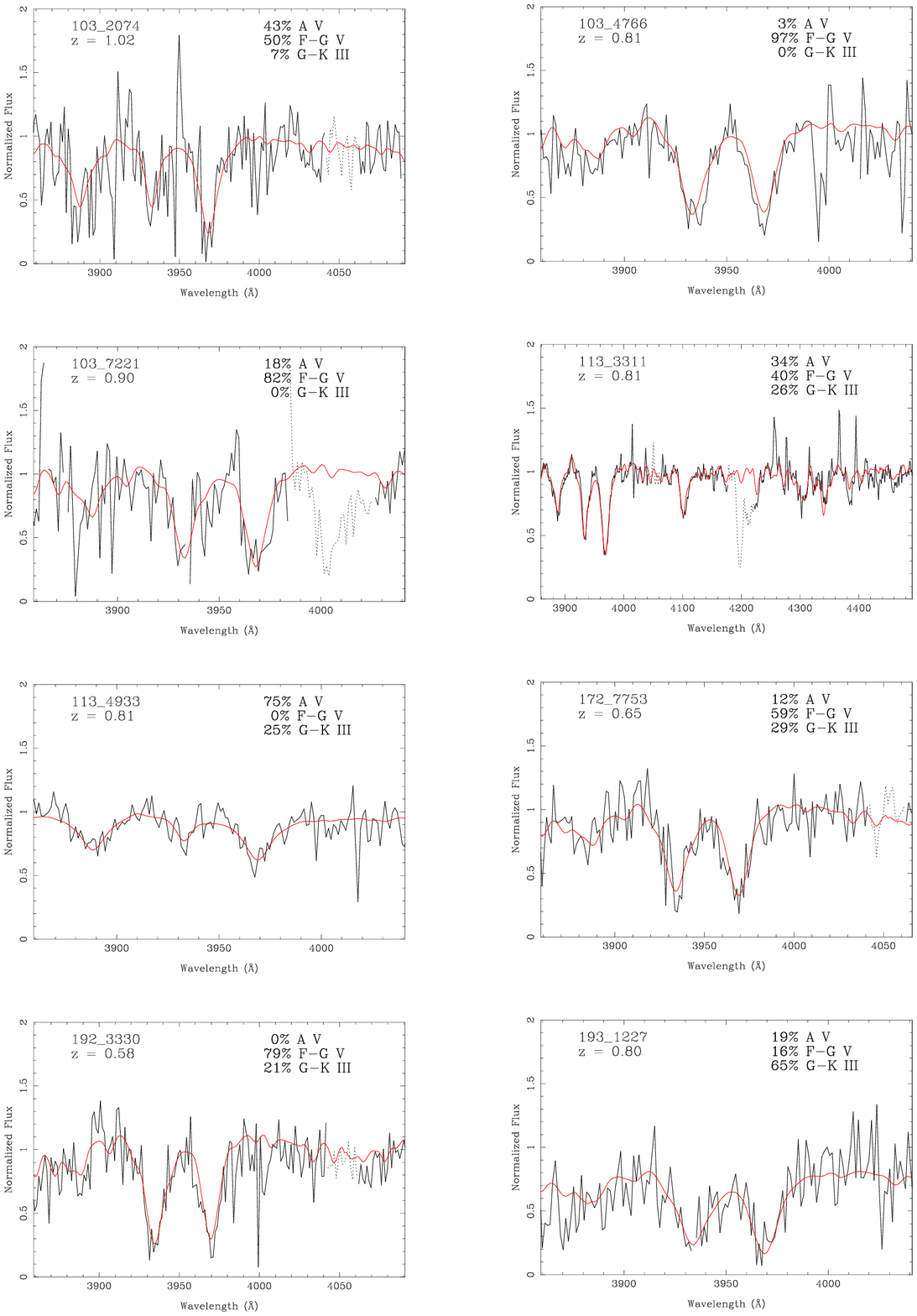,width=16cm,angle=0}}
\vskip -50pt
\figcaption{continued}
\end{figure*}

\begin{figure*}[t]
\figurenum{2}
\vskip -30pt
\centerline{\psfig{file=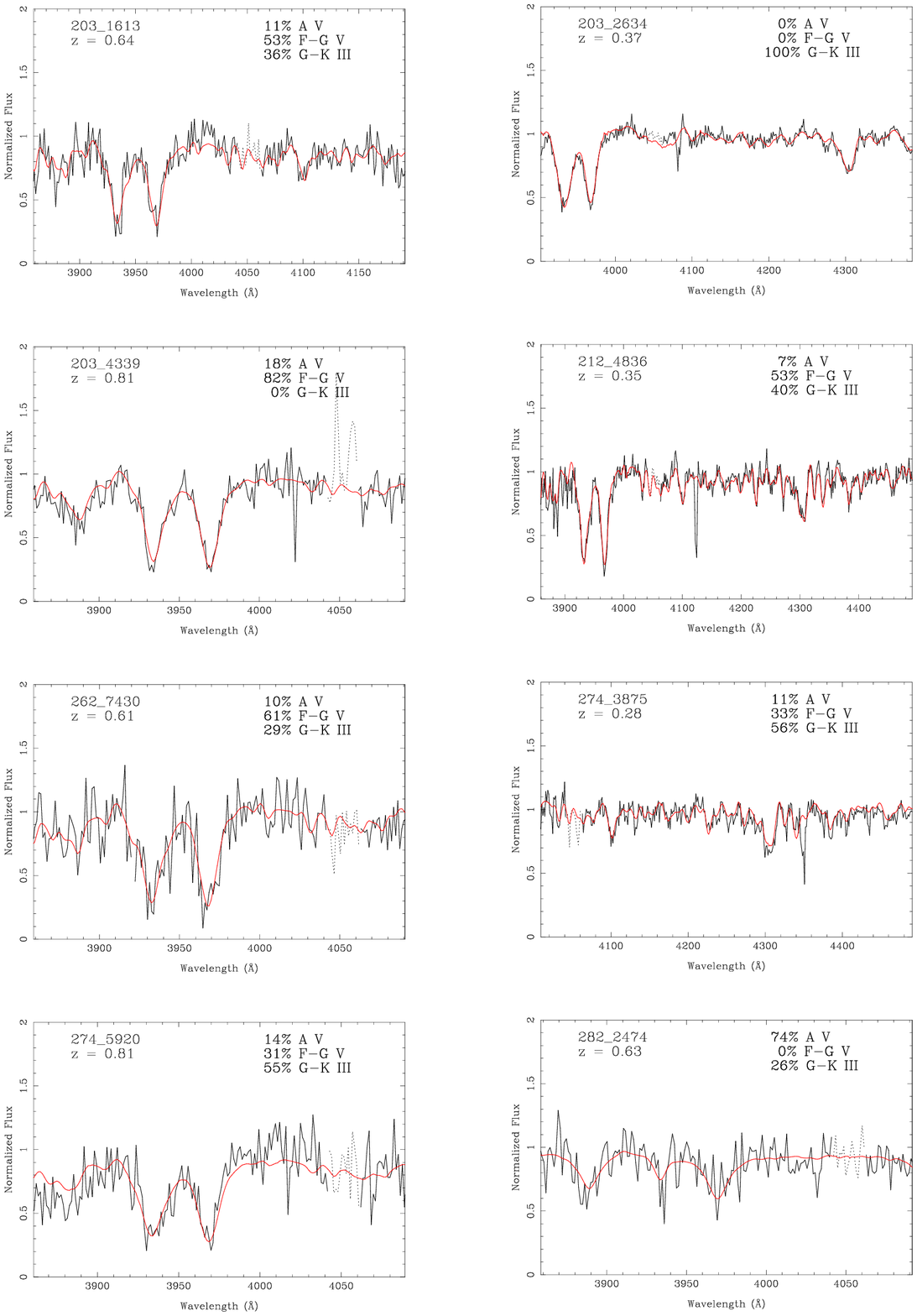,width=16cm,angle=0}}
\vskip -50pt
\figcaption{continued}
\end{figure*}

\begin{figure*}[t]
\figurenum{2}
\vskip -30pt
\centerline{\psfig{file=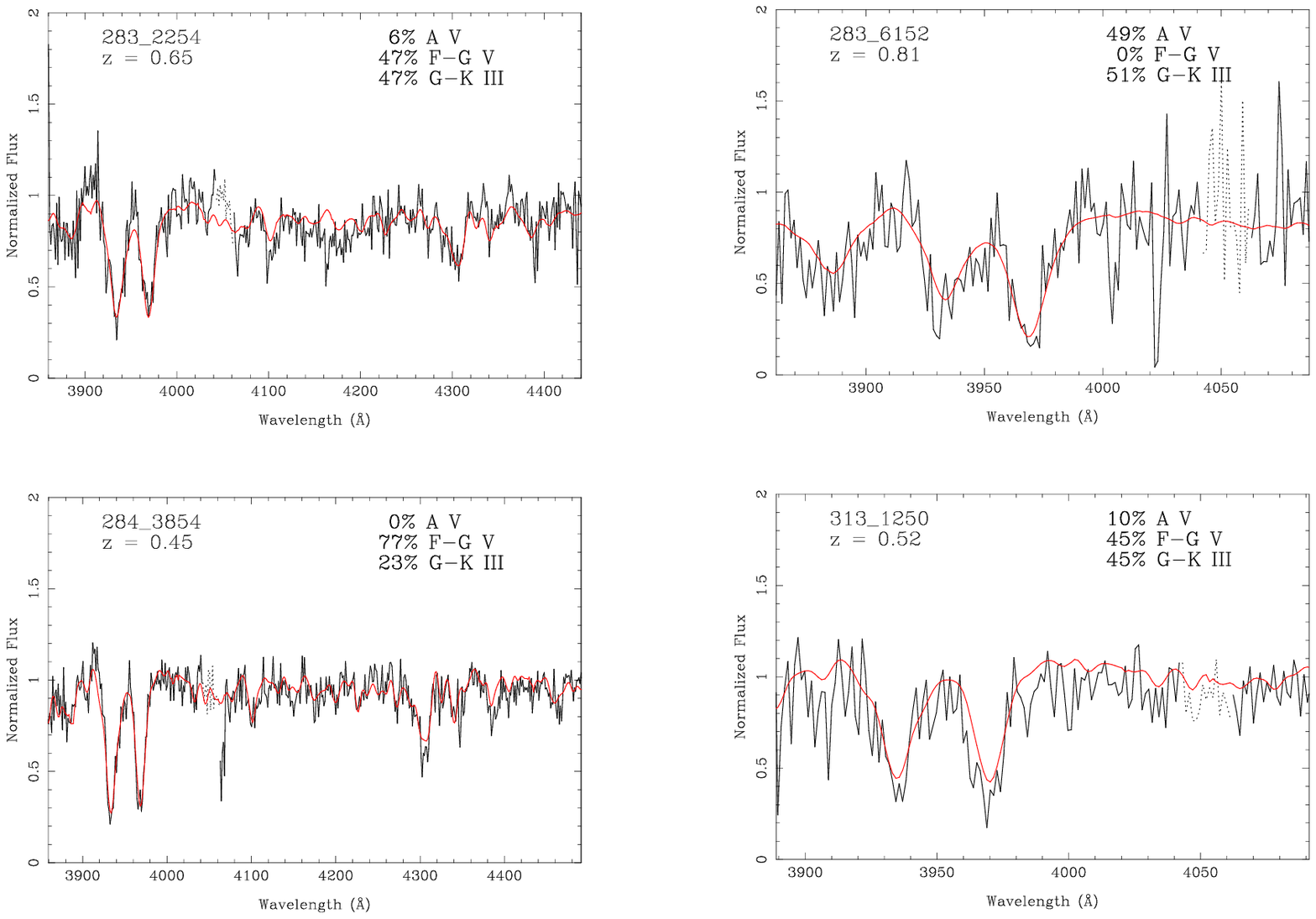,width=16cm,angle=0}}
\vskip -320pt
\figcaption{continued}
\end{figure*}


\subsection{Photometric Properties}

From the HST $V$ and $I$ images, we fit a two-dimensional disk/bulge
model to obtain the structural parameters.  The details of this fit
are in Simard~\etal\ (2002). Our galaxies span a range of types; we
place them on the Fundamental Plane by using half-light properties
instead of bulge-fit properties. All plots below use the half-light
radii and surface brightness. We discuss the changes when using the
bulge properties as well.

Local FP studies use pure R$^{1/4}$ fits to estimate the total light
and half-light radii. For the early-type galaxies in our sample, we
perform both disk/bulge and pure bulge fits. The two properties that
go into the FP are the surface brightness and half-light radius. The
root-mean-squared (RMS) difference in surface brightness between the
two fits is 0.4 magnitudes. However, the important property for the FP
is the change in surface brightness relative to the change in the half
light radius. By plotting these two parameters on orthogonal axes we
minimize the uncertainties due to different photometric models.
Kelson~\etal\ (2000a) discuss this effect in detail and find that
various photometric fits have little effect on the FP, since
differences are manifested by moving along the plane instead of
perpendicular to it. For our sample, the RMS differences perpendicular
to the FP using the two methods is only 0.02 magnitudes. This small
amount is larger than most of our individual uncertainties, and thus
is insignificant. Since the disk/bulge decomposition offer superior
fits, we only use those results for our analysis.

The two-dimensional fitting provides structural parameters defined
along the major axis; thus, half-light radii are along the semi-major
axes of the galaxies. Local samples generally use parameters defined
in circular annuli, or use geometric mean radii. In order to compare
properly to local samples, we must convert our major-axis radii and
surface-brightnesses to those defined according to circular
annuli. This conversion is particularly important for elongated
objects and those galaxies which have disk/bulge components that are
not at the same position angle, since in the latter case, major-axis
half-light radii are not well defined, whereas circular aperture radii
are. We integrate the disk/bulge model in circular annuli to the
radius that contains half of the light, setting the average effective
surface brightness, SB$_e$, equal to L$_{\rm total}/2\pi~r_h^2$. The
local samples for early-type galaxies come from J\o rgensen~\etal\
(1994), Faber~\etal\ (1989), and Bernardi~\etal\ (2003). These samples
use circular or geometric mean apertures. The difference between
geometric mean and circular radii are insignificant for our purposes.

We calculate K-corrections using an analytical formula derived from
local galaxy spectra. This estimate is discussed in the Appendix.  We
have compared these corrections to those determined from the stellar
population models of Bruzual \& Charlot (1998) and find little
difference. Since most of our galaxies are at redshifts between 0.7 to
1.0 and observed in $V$ and $I$, the K-corrections do not vary much in
terms of rest frame $U-B$ color; at $z$=0.8, the observed $V-I$
translates well into $U-B$.

Table~1 presents the basic data. It contains Groth ID (Col. 1),
redshift (Col. 2), spectroscopic exposure time (Col. 3), bulge
fraction in the $I$-band (Col. 4), apparent $I$ total Vega magnitude
(Col. 5), rest-frame absolute $B$ magnitude (Col. 6), observed $V-I$
(Col. 7), rest-frame $U-B$ (Col. 8), $I$-band half-light radius in
arcseconds and kpc (Cols. 9 and 10), average $B$-band surface
brightness inside of the half-light radius (Col. 11), velocity
dispersion (Col. 12), offset from Fundamental Plane in magnitudes
(Col. 13), photometric type from Im et al. (Col. 14), and RA and DEC
(Cols. 15 and 16). All rest-frame quantities use H$_0=70$,
$\Omega_{\rm m}=0.3$ and $\Omega_\Lambda=0.7$.

\subsection{Kinematic Properties}

We use combined, spatially-integrated line profiles to measure the
kinematics. In only a handful of cases are we able to measure
spatially resolved kinematics. The wavelength region used in the
analysis varies dramatically due to the redshift range of the galaxies
and the placement of the spectra relative to the night-sky lines,
which pose a particularly difficult problem for measuring
absorption-line kinematics. Table~2 and Figure~2 present the
wavelength region used for each galaxy. Locally, velocity dispersion
estimates rely on the G-band region ($\lambda$ 4300), Mg triplet lines
($\lambda$ 5167,5173,5184), and the Ca II triplet ($\lambda$
8498,8542,8662). At redshifts greater than 0.7, the G-band and Mg
lines are pushed into the night-sky forest, and the Ca triplet goes
into the infrared region. Since high-resolution multi-object
near-infrared spectrographs do not readily exist, we are forced to
rely on spectral regions where optical spectrographs are efficient,
i.e., the H+K region. This region historically has not been used much
for velocity dispersions since the lines are intrinsically broadened
and suffer from template mismatch more than in other spectral
regions. In addition, if the spectrum contains a significant amount of
A-stars, then the high stellar rotation in A-stars might complicate
the interpretation of the measured dispersion. However, the recent
study of Kobulnicky~\& Gebhardt (2000) demonstrates that, with care in
choosing stellar templates, the H+K region can be used to measure the
internal motions of the galaxy. Using a local sample of late-type
galaxies, this study compared the 20\% line widths measured from the
H+K region with those from HI velocity maps. To within 20\% of the HI
value, all galaxies in their sample compare favorably. The concordance
was even better for the earlier-type galaxies; the mix of stellar
populations in late-type galaxies complicates the analysis, compared
to the relatively uniform population in early-types. Because we are
studying mainly early-type galaxies, our situation is
promising. However, we still need to determine the template population
carefully for each galaxy (see below).

Obtaining internal kinematic information requires a deconvolution of
the observed galaxy spectrum with a representative set of template
stellar spectra. Both the deconvolution process and the template
library are significant issues.  For deconvolution, we treat each
spectrum using two different techniques: a maximum-penalized
likelihood (MPL) estimate that obtains a non-parametric line-of-sight
velocity distribution (LOSVD), and a maximum likelihood estimate of
the best-fit Gaussian LOSVD. Both estimates use a similar fitting
procedure. The programs proceed as follows: we choose an initial
velocity profile, either in bins for MPL or an analytic Gaussian
distribution. We convolve this profile with a weighted-averaged
template (discussed below) and calculate the residuals to the galaxy
spectrum. The program varies the velocity profile parameters---bin
heights or Gaussian $\sigma$'s---and the template weights to provide
the best match to the galaxy spectrum. The MPL technique is similar to
that used in Saha~\& Williams (1994) and Merritt (1997). However, the
difference is due to the simultaneous fitting of the velocity profile
and template weights. This fitting is essential since the template
profiles vary greatly in the H+K region, reflecting the relative
amount of old and young-star light. In fact, due to the large
variation among spectral types, the H+K region becomes an excellent
probe for disentangling the spectral make-up from the internal
kinematics. For the parametric estimate, we also include a
Gauss-Hermite polynomial fit (van~der~Marel~\& Franx 1993) using the
first four parameters---$h_3$ and $h_4$ (similar to skewness and
kurtosis)---in addition to the velocity centroid and dispersion. We
find no significant differences between the Gaussian fits compared to
the fits with additional moments, likely due to the low
signal-to-noise in the spectra. We report only the results for the
Gaussian fits.

The other important issue is the correct template. Ideally, one
observes the template library through the same set-up as the galaxy
data, as is generally done for local analysis.  However, for
redshifted data, it is not possible to obtain individual stellar
spectra covering the same rest-frame wavelengths and spectral
setup. We must use either synthetic spectra or spectra taken through
other spectroscopic configurations, and transform these to our
system. We use templates from two sources: the Coud\'e Feed Spectral
Library at KPNO (Leitherer~\etal\ 1996), and the stellar templates
observed by Kobulnicky~\& Gebhardt (2000). The KPNO library consists
of 684 stars having a broad range of spectral types, with 1.8~\AA\
FWHM resolution. We select a subsample of 21 stars representing A
main-sequence through K giant stars. This subsample provides good
overall type coverage without requiring an extensive number of
stars. The study of Kobulnicky~\& Gebhardt uses only 12
stars---covering a similar spectral range, but has higher resolution
of 1.2~\AA\ FWHM.

Template and galaxy spectra are continuum-divided. The continuum
division alleviates the need for accurate flux calibration. We use a
local continuum for the estimate as opposed to a global $n$-parameter
fit to the full spectrum. The continuum estimate relies on binned
windows from which we measure a robust mean (the biweight, see Beers,
Flynn, \& Gebhardt 1990) from the highest 1/3 of the points (in order
to exclude absorption lines), and then interpolate between these mean
values. We have tried a variety of continuum estimates (in terms of
number of local windows and number of points used in the local
averaging) and find insignificant differences in the kinematic
results.

We then transform each template to the resolution and redshift of each
galaxy spectrum. The resolution depends on the individual slitlet
(variation in the milling of the mask can cause slight variations of
the slit width and smoothness of cut) and also on the wavelength (in
particular, the ends of the wavelength coverage have worse
resolution); we measure the spectral resolution locally for each
galaxy in order to perform a proper kinematic analysis. The night-sky
lines provide the local resolution. At red wavelengths, the numerous
sky lines require multi-line fitting, with multi-line Gauss-Hermite
fits on spectral windows of 50~\AA\ each. The relative positions of
the lines were determined from high signal-to-noise, high-resolution
spectra (Osterbrock 1997). The fits produce the instrumental
resolution at each wavelength interval, and we choose that resolution
appropriate for the position of the usable features in the galaxy
spectrum. For the blue wavelengths (4000-6000 \AA), there are only a
few night-sky lines, forcing us to use an average resolution for the
entire range. The instrumental profile is well represented by a
Gaussian distribution plus a slightly negative tail component. We make
no correction for effects from either slit illumination or slit width;
however, since the seeing is typically larger than our slits and the
observed galaxy dispersions are larger than the instrumental width,
these effects are minimal. Having determined the appropriate spectral
resolution, we then redshift the template and convolve to this
resolution.  Redshifting the template causes the spectral resolution
to decrease by 1+$z$, and this must be included when calculating the
convolution. We convolve using a Gauss-Hermite polynomial with
$\sigma^2 = \sigma^2_{\rm g} - (\sigma_{\rm t}(1+z))^2$ and
$h_4=-0.07$. $\sigma_{\rm g}$ is the resolution for the galaxy
spectrum (measured from the sky lines), $\sigma_{\rm t}$ is that of
the templates, either 0.9~\AA\ for the KPNO library or 0.51~\AA\ for
the Kobulnicky~\& Gebhardt sample, and $h_4$ represent the negative
tail of LRIS. Since the template sample is larger for the KPNO
library, we prefer those. However, for the most distant galaxies, the
spectral resolution for the KPNO sample is too low---due the
broadening from 1+$z$---and we are forced to use the better resolution
Kobulnicky-Gebhardt sample. Table~2 and Figure 1 present the template
library used for each galaxy. Alternatively, we could broaden the
galaxy spectrum slightly to match the redshifted KPNO templates; we
prefer not to do this since many of the galaxies have small velocity
dispersions, and we want to optimize the signal. We have analyzed many
of the galaxies using both template libraries and find no significant
differences. Figure~2 shows the spectra for a sample of the galaxies,
and the Gauss-Hermite fit convolved with the template.

A handful of galaxies contain both the H+K region and the G-band
region. For these, we have measured the internal kinematics separately
over those regions. Figure~3 plots the dispersion measured from G-band
region (we used 4150\AA -- 4430\AA) versus H+K region (3850\AA --
4150\AA). Good agreement is obtained, with dispersions derived from
the H+K region on average only 4($\pm5$)\% smaller than the dispersion
measured in th G-band. Furthermore, Figure~3 contains galaxies where
the H+K dispersion estimate comes from the blue spectrum and the
G-band estimate is from the red spectrum. The concordance of the two
implies that the relative calibration of the instrumental profile is
correct.

We use Monte Carlo simulations to measure the uncertainties on the
dispersions. For each realization, we generate a simulated galaxy
spectrum based on the best-fit velocity profile and an estimate of the
root-mean-square (RMS) of the initial fit. Table~2 presents the
signal-to-noise per pixel for each of the galaxies. The initial galaxy
spectrum comes from the best-fitting template convolved with the
measured LOSVD. This provides a galaxy spectrum with essentially zero
noise (the noise in the template is insignificant), plotted as the red
smooth line in Figure.~2. From that initial spectrum, we then generate
100 realizations and determine the velocity profile, and hence the
velocity dispersion, each time. Each realization contains flux values
at each wavelength position randomly chosen from a Gaussian
distribution, with the mean given by the initial galaxy spectrum and
the standard deviation given by the RMS of the initial fit. The 100
realizations of the velocity profiles provide a distribution of values
from which we estimate the 68\% confidence bands.


\psfig{file=gebhardt.fig3.ps,width=8.8cm,angle=0}
\figcaption[sig.ps]{The velocity dispersion in \kms\
measured from the H+K lines relative to the G-band estimate.
\label{fig3}}
\vskip 0.3cm


Besides the dispersion uncertainty, an important aspect of the
simulations is to determine the measured bias. Often, and especially
in low S/N data, the measurement technique itself may bias the
parameter value. The best way to study this is to examine the
distribution of values in the Monte Carlo simulation. In every case,
the bias in the dispersion is much smaller than the uncertainty and
the average bias is zero. The Monte Carlo procedure incorporates the
major sources of uncertainty in this analysis, namely, the broadened
lines of the H+K region, the variable template mix, and low S/N. Thus
we are confident that our dispersion estimates are
unbiased. Furthermore, we have run all following analyses using a
strict cut of better than 15\% accuracy on the dispersion
measurements. This cut leaves 14 galaxies in the sample, and we find
no differences in the results versus using the full sample.

The relative uncertainty for the velocity dispersion scales roughly
linearly as the S/N. A S/N of 5 per pixel (the lowest in the sample)
implies an uncertainty around 40\% and a S/N of 25 per pixel (the
highest) implies an uncertainty around 5\%.

We have one galaxy in common with a previous measurement. The
quad-lens galaxy, 093\_2470 (HST14176+522), has a measured dispersion
of $230\pm14$ km/s by Ohyama et al. (2002). Our measured value is
$202\pm9$, which is barely consistent with the previous measurement.
This galaxy spectrum has the highest signal-to-noise ratio in our
whole sample, and thus the best measured dispersion. We have
thoroughly checked systematic issues by changing the spectral fitting
range (H+K only, G-band region only, and full spectrum) and the used
templates. We find differences consistent within our quoted
uncertainty.  Given that our spectrum has higher S/N than that of
Ohyama et al., we choose to use our measurement.


\vskip 0.3cm 
\begin{figure*}[t]
\centerline{\psfig{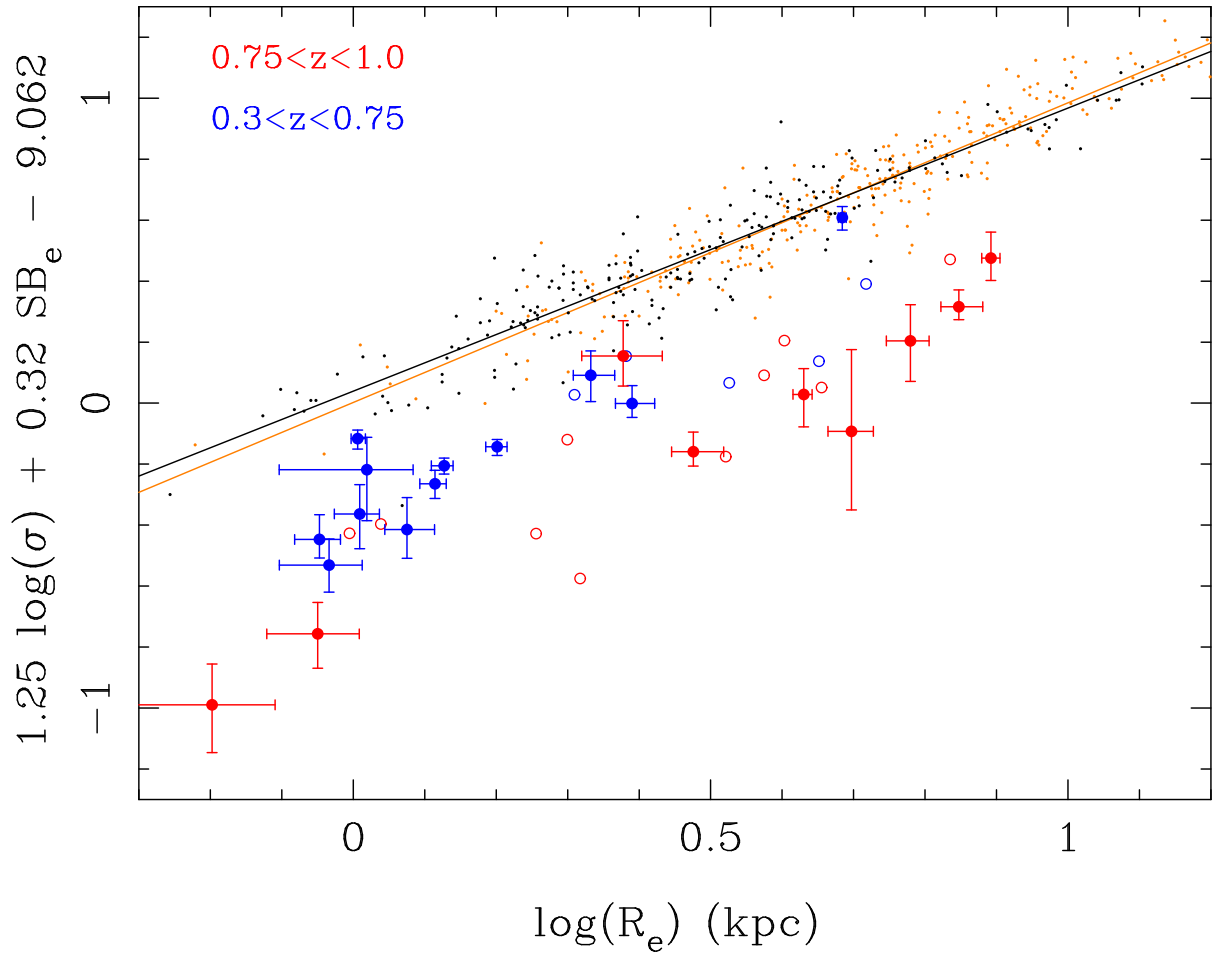}}
\figcaption[sig.ps]{The Fundamental Plane for early-type galaxies. The
red points represent field galaxies with redshifts greater than 0.75,
and the blue points are those below 0.75. The small points represent
the local sample, and the lines are linear fits to each sample, with
black representing J\o rgensen~\etal\ (1994) and orange Faber~\etal\
(1989). The solid points with error bars are the early-type galaxies
and the open circles are disk-dominated galaxies. SB$_e$ is the
average surface brightness within $r_e$ in units of rest-frame $B$
magnitudes. $\sigma$'s are aperture corrected. We use the fit of
Bernardi~\etal\ (2003) to measure the magnitude offset from the
high-$z$ sample.
\label{fig4}}
\end{figure*}
\vskip 0.3cm


\subsection{Aperture Correction}

Since the velocity dispersion may vary with radius in galaxies, and
since we are using spectra extracted from different physical sizes, we
must correct the velocity dispersion to a common radius. The local
analysis of J\o rgensen~\etal\ (1994) corrects to an aperture of
radius 3.4\arcsec\ at the Coma distance, and we adopt that value. We
determine the aperture correction for each galaxy individually using
its brightness profile and redshift. However, the radial profile for
the velocity dispersion is unknown, and we have to approximate this
variation. We use the local samples of Gonz\'{a}lez (1993) and Ziegler
\& Bender (1997) to determine a characteristic gradient in the
dispersion profile. Gonz\'{a}lez's data extend from r=0--R$_e$, while
Ziegler~\& Bender's data extend much further out to 3~R$_e$. The
variation seen inside R$_e$ is the same, namely, the variation is
represented by a power-law with exponents ranging from --0.1 to 0. At
radii beyond R$_e$ the profile becomes shallower, represented by
log~$(\sigma) = -0.1 (r/{\rm R}_e)^{0.75}$ (Ziegler \& Bender 1997),
but there is a large variation of the slope. There appears to be no
obvious trend between the dispersion profile with any other galaxy
property, so we must use average values for our sample.

The radial extent of the extracted galaxy spectrum depends on the
half-light radius, the seeing disk, and the slit width. We use an
extraction window that is equal to the half-light radius convolved
with the seeing disk; this procedure optimizes the signal while
attempting to achieve similar extraction windows for the full sample.
The data contain galaxies with half-light radii that range from
0.1\arcsec\ to 1\arcsec, and the slit width and seeing are about an
arcsecond; we then have extracted spectra that contain galaxy light
with aperture radii that vary from 1 to 5~R$_e$. Using this range of
extraction radii and the above estimates for the shape of the
dispersion profile, we model each galaxy as an R$^{1/4}$
surface-brightness profile to estimate the range of aperture
corrections with circular apertures. Using Ziegler~\& Bender's
$\sigma$ profile, we find that the average dispersion correction is
+10\%, and a range of power-law exponents from 0 to --0.1 yields
corrections of 0\% to +15\%, respectively. Gebhardt~\etal\ (2000)
present dispersions inside various radii for a sample of 13 nearby
ellipticals. They also find no obvious trends amongst galaxy
properties; however, their data imply that the aperture dispersion
variations are even smaller, with corrections ranging from --10\% to
+10\% and averaging near zero. We compromise and increase all
dispersion estimates by 10\%.

\section{Fundamental Plane}

Table~1 includes the galaxy properties used in the Fundamental
Plane. We report the measured velocity dispersion and not the aperture
corrected value; as mentioned above, the latter is obtained by
increasing the first by 10\%. The surface brightness---using circular
annuli---is in rest-frame $B$ magnitudes (K-corrections are discussed
in the Appendix). The following discussion in Section 3 and 4 is
limited only to early types as classified by Im et al. (2002). Section
5 considers the disk-dominated galaxies as well.

\subsection{Comparison with Local Samples}

Faber~\etal\ (1989), J\o rgensen~\etal\ (1994), and Bernardi~\etal\
(2003) provide the local calibration for the Fundamental Plane. The
Faber~\etal\ sample is measured in the B-band, J\o rgensen~\etal\ use
Gunn r-band, and Bernardi~\etal\ use the Sloan Gunn filters ($g^*,
r^*, i^*, z^*$). Since at $z\approx 1$ the observed $V$ and $I$
correspond to restframe $U$ and $B$, the most direct comparison is to
$B$-band derived quantities. This is especially important since the
slope of the plane may depend on band (e.g., Bernardi~\etal\ find a
linear trend of the slope with band; the slope increases slightly for
redder bandpasses). Thus, we utilize the $B$-band Fundamental Plane,
as originally used in Faber~\etal\

Figure~4 plots the edge-on view of the Fundamental Plane. Two samples
of nearby galaxies are shown: those from Faber~\etal\ and J\o rgensen
et al. We do not include the galaxies from Bernardi~\etal\ since they
are not public. The two local samples plotted in Figure~4 provide a
consistency check. The main differences are that the Faber~\etal\
sample is a magnitude-limited sample of both field and cluster
ellipticals, whereas J\o rgensen~\etal\ concentrate mainly on cluster
galaxies and include many S0s. Thus, the Faber sample has little bias
in terms of defining a ``typical'' local elliptical, while the J\o
rgensen sample includes fainter galaxies though not necessarily
ellipticals. Although many objects are in common, we plot both local
samples in full to show the distribution of values. We use an average
offset to transform the Gunn $r$ colors of J\o rgensen~\etal\ to the
$B$ magnitude system used by Faber~\etal\ ($B-r$=1.25). As expected,
since J\o rgensen~\etal\ go fainter, they tend to have more smaller
galaxies, whereas the Faber~\etal\ sample is weighted towards large
ellipticals. Both samples provide a similar FP relation, with
log~$(R_e) \approx 1.25~{\rm log}~\sigma + 0.32~{\rm SB}_e - 9.062$,
where we have minimized the residuals in the orthogonal direction.

Bernardi~\etal\ (2003) also measure a tight FP relation from their
8000 galaxies. Their slope is higher than the one used here (1.49
compared to 1.25). Since the differences in the Bernardi~\etal\ slopes
in their four bandpasses are small, the $B$-band FP slope derived from
their data should not be much smaller than 1.49. We have tested
whether our results are affected by the choice of FP parameters, using
a variety of slopes (from 1.1 to 1.6) and find little difference. The
main reason for this is that the large offsets in luminosity evolution
that we see are much larger than the subtle differences caused by
using a different slope for the FP.

We divide the distant sample in Figure.~4 into low ($z<0.75$) and high
($z>0.75$) redshift. In both the low and high $z$ datasets we see a
tight correlation in the FP. There is a trend for the more distant
galaxies to be further from the local relation, as expected from
evolution. Figure~5 plots the offset in magnitudes from the
local relation as a function of redshift. We assume for this plot that
all of the evolution in the FP is due solely to stellar luminosity
evolution. The solid black line in the plot is a cubic fit to the
data, forcing the local universe to have zero shift.

Included in Fig.~5 are magnitude offsets for the individual galaxies
in the distant cluster MS1054-03 at $z=0.83$ (blue dots). Our
photometry has been run on the van~Dokkum~\etal\ galaxies using the
GIM2D software (Simard 2002) to make sure that we are on the same
system. The mean magnitude offset for the cluster in Fig.~5 comes from
van~Dokkum et al. 1998, and the individual points come from our re-analysis
of their data. We find no systematic offset. Furthermore, the two E+A
galaxies in MS1054-03 lie along the same relation as we find for the
field galaxies at that redshift. Van~Dokkum~\etal\ concluded that
these two galaxies were recent acquisitions to the cluster, and thus
are likely field galaxies. In addition, J\o rgensen~\etal\ (1999)
present detailed analysis of the evolution of the Fundamental Plane
for cluster galaxies spanning $0<z<0.6$. Figure~5 includes surface
brightness offsets from these distant clusters. An offset between the
two relationships clearly exists; field galaxies show significantly
more brightening than cluster galaxies. This result suggests that
field galaxies have a more recent formation epoch than cluster
galaxies.

Im~\etal\ (2002), using the early-type galaxy luminosity function at
high redshift, find a significant brightening in L$^*$ for field
galaxies as well. The results presented here are consistent; in Fig.~5
we include the two measurements from Im~\etal\ (black circles with
error bars). We conclude that there is about a 1.6 magnitude
brightening around z=0.8 and 2.4 around a redshift of 1. Other studies
have found similar magnitude offsets. Cohen (2002), using the
luminosity function, find approximately two magnitudes of brightening
in their field samples at redshift near 1. Koo~\etal\ (2003), using a
bulge sample, find an average luminosity brightening of 1.5 magnitudes
in the redshift range $0.72<z<1.1$; if we use the same redshift
interval, our average offset is 1.75 magnitudes. At redshifts below
0.4, Treu~\etal\ (2002) and van~Dokkum~\etal\ (2001) find consistent
results from their Fundamental Plane analyses that are consistent with
our low-redshift sample.


\vskip 0.3cm \psfig{file=gebhardt.fig5.ps,width=8.8cm,angle=0}
\figcaption[dmu.ps]{Residuals from the local Fundamental Plane
assuming that all evolution is due to stellar luminosity
evolution. Filled green circles represent early-type galaxies, and
open green circles are disk galaxies. The blue squares are mean
cluster values from the compilation in J\o rgensen~\etal\ (1999), and
the blue circles are values for individual galaxies in the distant
cluster MS1054-03 at $z=0.83$ (van~Dokkum et al. 1998, see text).  The
black circles with error bars are mean field values reported in
Im~\etal\ (2002) using evolution in L$^*$. The solid black line is a
cubic fit to the field points. The red line and blue lines are
predicted offsets from stellar population models with formation times
$z_f=1.3$ (for the field) and $z_f> 3$ (for the clusters),
respectively.
\label{fig5}}
\vskip 0.3cm


An obvious potential bias is whether the large magnitude brightening
results from a S/N limitation; i.e., the magnitude brightening that we
see could result if only bright galaxies make it into the
sample. However, this bias does not appear to be significant. First,
we are considering magnitude offsets from the FP and not actual
magnitudes. Thus, two galaxies can have the same offset but have very
different effective surface brightnesses; for the three galaxies near
$z=1$, their surface brightnesses range by 3.5 magnitudes yet they all
show nearly identical magnitude brightening.  Second, we do measure
some galaxies with much less brightening---in particular, 193\_1227 at
$z=0.8$---which demonstrates that any such galaxies in the sample are
included. It appears that we are not measuring an outer envelope of a
distribution but instead its actual extent.

Bernardi~\etal\ (2003) estimate this bias in the SDSS sample using
Monte Carlo simulations of their data. They find that they have a 0.1
magnitude bias due to their magnitude-limited sample. The galaxies
with the smallest magnitude offsets are not in their sample since they
are too dim, and thus biasing the fit to the magnitude offset for the
full sample. Unfortunately our sample is not as uniform as the
Bernardi~\etal\ sample and we cannot easily simulate datasets that
mimic our selection criteria (for a complete discussion of the
selection criteria see Simard~\etal\ 2003 for the photometric
selection and Weiner~\etal\ 2003 for the spectroscopic selection). For
example, the galaxies in this sample have exposure times ranging from
35 to 310 minutes. As explained before, the selection criteria for the
FP sample includes adequate S/N and appropriate spectral coverage that
includes usable absorption lines. Thus, our best way to study
potential biases is to examine correlations with S/N. Figure~6 plots
the S/N per \AA\ for the full sample as a function of magnitude offset
from the local FP. We break the sample into low-$z$ ($z<0.7$) and
high-$z$ ($z>0.7$), since Fig.~5 suggests a strong trend with
redshift. If we have a bias such that only galaxies with large
luminosity evolution make it into our sample, then as we include
fainter galaxies we might expect a smaller amount of luminosity
evolution for them. This is not the case in Fig.~6. If we divide the
sample into two groups based on S/N, the high-$z$ galaxies with
S/N$<10$ have, on average, the same amount of luminosity evolution as
the galaxies with better S/N. We can also plot in Fig.~5 only the high
S/N galaxies; the best-fit trend with redshift found using high S/N
galaxies is indistinguishable from the trend using the full sample.
Thus, we are confident that the trend and scatter seen in Fig.~5 are
real.


\vskip 0.3cm \psfig{file=gebhardt.fig6.ps,width=8.8cm,angle=0}
\figcaption[gebhardt.fig6.ps]{Residual from the local Fundamental
Plane versus the signal-to-noise (S/N) per \AA\ of the spectrum.  The
open circles are those galaxies with $z<0.7$ and the filled circles
are those with $z>0.7$. As seen in Fig.~5, the high-$z$ objects have
larger luminosity evolution. However, among the high-$z$ sample there
appears to be no trend with S/N.
\label{fig6}}
\vskip 0.3cm


\subsection{Comparison with Previous Work}

Several papers have measured FP offsets for field galaxies. The most
recent include van~Dokkum \& Ellis (2003), Treu~\etal\ (2002), van
Dokkum~\etal\ (2001), and Koopmans \& Treu (2002). In Figure~7, we
repeat the points in Figure~5 but now include these four additional
samples. Treu~\etal\ (2002) study FP evolution from 30 field galaxies
out to a redshift of 0.6. They provide averages in five redshift bins,
which we will use for comparison. Van~Dokkum~\etal\ (2001) present 18
field galaxies out to a redshift of 0.55. They provide averages in
three redshift bins. Van~Dokkum \& Ellis (2003) provide nine field
galaxies selected from the Hubble Deep Field out to redshift
1.0. Koopmans \& Treu (2002) provide one galaxy at redshift 1.0.

The Treu~\etal\ points fall along the offsets in our sample. The
van~Dokkum et al. (2001) points follow a similar trend, but with
slightly less luminosity evolution. The van~Dokkum \& Ellis (2003)
points are significantly different from our field sample, and quite
similar to the cluster galaxy offsets. The single galaxy of Koopmans
\& Treu at $z=1$ is more consistent with the cluster galaxy offsets
and the authors indeed suspect that this galaxy may be in a
proto-cluster. The most important aspect of this plot is that there
are results from three independent groups, using different
observations and analyses. The agreement between our field sample and
that of Treu~\etal\ argues that there are no biases between our
groups. The field galaxies of van~Dokkum~\etal\ show less evolution
but still more than the cluster galaxies. In fact, van~Dokkum~\etal\
report an offset between their cluster and field sample of 0.2
magnitudes. Using their median redshift bin of 0.46, we find an
difference around 0.3 magnitudes. While this difference is small, it
translates into a large difference at redshift of one. Thus, even
though the comparison between our sample and that of van~Dokkum~\etal\
appears to be insignificant, the consequences extrapolating to
redshifts near one are important.

It is important to obtain larger samples of field galaxies below
redshift 0.7. Bernardi~\etal\ (2003) look for evolution in the FP
using 9000 galaxies from Sloan out to redshift 0.26. They find that
after correcting for selection effects, the luminosity evolution is
measured to be, at most, 0.1 magnitudes at $z=0.2$. Since we have no
data at $z=0.2$, we cannot compare directly, but this amount of
evolution is consistent with the numbers that we measure for higher
redshifts.


\vskip 0.3cm \psfig{file=gebhardt.fig7.ps,width=8.8cm,angle=0}
\figcaption[dmu2.ps]{Comparison of our residuals with those from other
sources. The small symbols are the same as in Fig.~5; solid triangles
come from redshift averages of Treu~\etal\ (2002), the open triangle
from the single galaxy of Koopmans \& Treu (2002) at $z=1$. The filled
black stars from redshift averages in van~Dokkum~\etal\ (2001), and
the filled red stars are the individual galaxies in van~Dokkum \&
Ellis (2003). We have transformed all values to the same cosmology of
H$_0=70$, $\Omega_{\rm m}=0.3$ and $\Omega_\Lambda=0.7$.
\label{fig7}}
\vskip 0.3cm


Fig.~7 shows that our distant field sample continues the trend seen in
other, lower-redshift, field samples. The magnitude brightening
deduced from our sample is in agreement with the extrapolation from
Treu et al. (2002). At redshifts below 0.6, the agreement between our
data and Treu et al. is excellent. The field sample of van Dokkum et
al. (2002) is not in as good agreement, but the differences are not
large given the uncertainties.

The biggest difference is between our field sample and that of
van~Dokkum and Ellis (2003). We have excluded differences in the
fitting techniques for both photometry and kinematics. For the
photometry, we have run independent fits and find very similar
magnitude offsets; the mean difference between in magnitude offset
between our photometry (Simard et al. 2002) and theirs (van~Dokkum \&
Franx 1996) is consistent with zero, with a scatter smaller than 0.1
magnitudes.  For the kinematics, we have observed one galaxy in common
with the sample of van~Dokkum \& Ellis, using the same observational
setup as our other galaxies. For this galaxy, HDFN\_J
$123700.56+621234.7$ at $z=0.562$, they measure a velocity dispersion
of $114\pm19$ and we measure $111\pm10$. At least for this galaxy,
both groups find a similar offset. Ironically, the offset for this
galaxy lies exactly in between the average offsets for our sample and
van~Dokkum \& Ellis (2003). Thus, there are no obvious reasons for the
differences from the analysis techniques. We conclude that the major
difference is likely due to sample selection. In fact, van~Dokkum \&
Franx (2001) argue that their FP sample is likely biased due to what
they call ``progenitor bias'', causing their galaxies to not be a fair
representation of galaxies today. The main effect of the ``progenitor
bias'' would be to cause the observed FP magnitude offset to be less
than it actually is since one would be looking at the reddest, most
evolved systems. Our selection criteria is mainly based on bulge
fraction and therefore may be a more fair representation of local
ellipticals. Two of the nine galaxies in van~Dokkum \& Ellis (2003)
have similar magnitude offsets as we measure. Including the van Dokkum
\& Ellis objects with ours would substantially increase the intrinsic
scatter of the FP. Clearly, studies with larger samples and
well-defined selection effects are needed.

\section{Integrated Color Distributions}

Colors provide additional insight into galaxy evolutionary history,
since changes in the stellar population affect both total brightness
and color. However, it is difficult to compare colors of galaxies
observed at different redshifts since all data must be transformed to
a common photometric system, requiring accurate spectral energy
distributions (SED's) for all galaxy types. Furthermore, at redshifts
beyond $z=0.8$, the observed $V$ represents rest-frame $U$ magnitude,
for which we do not have large local samples for comparison. Thus to
use color information, the two main obstacles are obtaining both
accurate transformations to a local rest-frame color and an accurate
local galaxy color distribution.

Accurate color transformations are best determined using actual galaxy
spectra. Spectral synthesis models do not reflect the wide variation
seen in local galaxies, and specifically they do not include dust
effects. Since our observed bands are $V$ and $I$ at redshifts from
0.3 to 1, we must include UV, as well as optical, spectra for the
local samples. In the Appendix we describe rest-frame color estimates
and K-corrections.

The local galaxy $U-B$ color distribution comes from the compilation
by Prugniel \& Heraudeau (1998), available from the CDS
(http://cdsweb.u-strasbg.fr/CDS.html). This database consists of
observations from a variety of sources. From approximately 5000
galaxies in the compilation, we choose those with T-type less than 0
and $U-B$ uncertainty less than 0.1 magnitude, yielding 290 early-type
galaxies. Additionally, since the color of the galaxy depends on its
magnitude, we must check that the absolute magnitudes of both samples
(distant and local) are consistent. The absolute magnitude range for
the distant early-type sample is $-22.5<{\rm M}_{\rm B}<-19.4$;
however the proper local comparison should include the magnitude
dimming. Thus, the range that we select from the local sample is
$-21.5<{\rm M}_{\rm B}<-18.7$. This cut leaves 219 local
galaxies. Including the brighter objects would not change the results
significantly since the local $U-B$ distribution for all early-types
is quite narrow. The mean $U-B$ for local early-types is 0.52, using
either the full sample or the magnitude cut. Using an extreme cut on
T-type less than --3 produces only a slightly redder $U-B$ of
0.55. Another important point is to consider possible differences
between field and cluster galaxies, since our sample comes from the
field. Again, the distribution of $U-B$ is so small that any
differences between local field and cluster galaxies will be
insignificant for our purposes. Once larger samples are studied, both
distant and local, this comparison would be worthwhile.


\vskip 0.3cm \psfig{file=gebhardt.fig8.ps,width=8.8cm,angle=0}
\figcaption[ub.ps]{Rest frame integrated $U-B$ colors for the distant
early-type field sample (blue solid line), bulge component of the
early-type field sample (blue dotted line), MS1054-03 cluster (green
line), local sample (black line), and the Koo et al. (2003) bulge
sample from the distant field (red dotted line). The uncertainties for
colors of the distant field sample is on average 0.03 mags, which is
significantly smaller than the width of the color distribution.
\label{fig8}}
\vskip 0.3cm


Figure~8 compares $U-B$ distributions for total galaxy light in our
distant early-type field sample, their bulge components, the cluster
MS1054-03 at $z=0.83$ from van~Dokkum~\etal\ (1999), a bulge field
sample from Koo~\etal\ (2003), and the local sample (total
light). Each curve represents an adaptive kernel density estimate
(Silverman 1986).  The details of the adaptive kernel are given in
Gebhardt~\etal\ (1996). The goal is to measure a smooth underlying
density distribution from a small number of points. A histogram
represents the most basic and coarse approximation to a density
distribution. A more robust estimate relies on varying the window size
depending on the local density; this is the adaptive kernel. We use an
initial fixed kernel width to estimate the local density, and then run
a kernel which varies in width inversely according to that initial
density to obtain the final density. This estimate preserves the
information content in low-density relative to high-density regions.

For the total-galaxy light samples, the trends in Fig.~8 show that the
bluest objects are the field galaxies at high-z, the next are the
cluster galaxies at z=0.83, and the reddest sample contains the local
galaxies. These trends are consistent with passive evolution and
having the cluster ellipticals that existed in MS1054 at $z=0.8$ form
stars earlier than our field early-types at the same epoch. As
expected, the bulge components for the field galaxies are redder than
their total colors and are consistent with the bulge colors of the
Koo~\etal\ bulge sample. These color distributions are discussed in
detail below and in Koo~\etal. The Koo~\etal\ bulge sample consists of
all galaxies with bulges brighter than $I=23.6$ and includes 44\% of
our galaxies---thus, the close agreement between the $U-B$ colors of
the two samples seen in Fig.~8 (0.04 mag) is expected.

\section{Discussion}

Viable evolutionary scenarios for early-type galaxies must match both
the changes seen in Fundamental Plane surface brightness and in
color. Complications arise due to the uncertainty in knowing which
galaxy property in the FP is changing with time. Another source of
confusion is that dust properties may change with time, and their
effects should be included.

Both size and velocity dispersion may evolve and affect the
distribution in the Fundamental Plane. A full treatment of the size
and dispersion distribution requires detailed modeling of the
selection functions and will not be explored here. However, the most
straightforward interpretation of luminosity evolution measured by Im
et al. (2002) is that of stellar-population evolution
only. Furthermore, preliminary analysis using a similar study to that
in Im et al. for the luminosity evolution shows little evolution in
the size function. There appears to be slight evolution in the
velocity dispersion---of order 10\% near z=0.8 but with large
uncertainty. This dispersion offset implies a change of only 0.16
magnitudes in surface brightness. Thus, the true luminosity evolution
may be 0.16 magnitudes less than the magnitude offset seen in Fig.~5
at $z=0.8$.

On theoretical grounds, hierarchical clustering predicts that mergers
will occur primarily from parabolic orbits. Homologous merger models
for such cases predict very little change in sigma but a doubling of
the radius (for equal-mass mergers). In general, radius will grow in
proportion to mass, with sigma staying constant. Since local galaxies
show that surface brightness goes inversely with radius, the effect of
mergers on the Fundamental Plane is to move galaxies along rather than
away from it. Thus, as the simplest approach, we consider FP evolution
to be due solely to luminosity evolution with no dust evolution, and
discuss possible consequences of this assumption.

\subsection{Magnitude Evolution}

The scatter in the FP and how it evolves over time is an important
quantity to measure. To measure the scatter, we first remove the
evolutionary effect by using the cubic fit in Fig.~5, where we have
assumed that all evolution is due to stellar luminosity
evolution. Figure~9 plots the Fundamental Plane corrected empirically
for this evolution. The correction uses a cubic in redshift given by:
$<{\rm SB}_e>_0 = <{\rm SB}_e>_z + 3.38 z - 4.97 z^2 + 4.011 z^3$. The
resulting root-mean squared (RMS) scatter orthogonal to the local
calibration in Fig.~9 is 7.8\% using only the early-type
galaxies. This scatter is slightly larger than that seen in local
samples: the Bernardi et al. (2003) sample has a scatter of 5.6\% in
residuals calculated orthogonal to the best-fit line. The Faber~\etal\
and J\o rgensen~\etal\ sample have similarly small scatter. Since the
scatter seen in our high redshift data is similar to that seen
locally, we are confident that measurement uncertainty does not hinder
our ability to probe evolutionary effects using the FP. Surprisingly,
the expected scatter due to the observational errors alone is around
7.5\%, almost equal to the measured RMS. If we include the intrinsic
scatter as measured from Bernardi et al., we would expect a measured
RMS of 9\%. The small scatter that we see in our sample is similar to
the scatter that Treu~\etal\ (2002) find in their sample of 30 field
ellipticals at lower redshifts. Given that the measured scatter is
nearly consistent with the expected scatter based on local samples, we
conclude that the intrinsic scatter at high redshift must be small.

Figure 9 includes both spheroidal and disk galaxies. Since most FP
analyses are done using a sample of only early-type galaxies, we must
be careful about introducing potential biases. Kelson~\etal\ (2000a)
present detailed analysis of this bias in their cluster at redshift
$z=0.33$. Comparing the FP derived from their 30 E and S0 galaxies to
that derived from their 22 S0/a and later type galaxies, they find
both that the FP scatter increases by a factor of two and the
later-type galaxies are systematically brighter.  Due to a strong
correlation between FP offset and velocity dispersion, they conclude
that most of this difference is due to stellar population
differences. Correcting for this age effect makes the spirals fall
onto the FP for their ellipticals, although still with a factor of two
increased scatter.

The signal-to-noise of our images, spatial resolution of galaxies at
redshifts near 0.8, and observed bandpass shifting (restframe $U$ and
$B$) make it difficult to rely on visual classifications as Kelson et
al. (2000a) have done. We rely instead on bulge/disk decompositions to
classify galaxies into early and late types (see Im~\etal\ 2002 and
Simard~\etal\ 2002). We also do not have enough late-type galaxies in
our sample to compare at the same level of detail used in Kelson et
al., but we do see a general trend of increased scatter from the late
types compared to the early types.  The green points in Fig.~9 have
bulge fractions less than 0.4 which we consider to be disk-dominated
galaxies; indeed, these galaxies often show obvious signs of spiral
structure in Fig.~1. These galaxies lie slightly above the line in
Fig. 9. This is expected since the increased star formation rate in
spirals tends to create a more constant surface brightness over time,
and hence less evolution. The difference between the late and early
types is about 0.4 magnitudes, where the late types have a smaller
offset at a given redshift. Inspection of galaxies with high bulge
fractions (those expected to be true ellipticals) shows no trends away
from the full population. This small difference is also seen in local
samples.  Falc\'{o}n-Barroso~\etal\ (2002) place S0--Sbc galaxies on
the Fundamental Plane defined for nearby cluster ellipticals. They
find that later type galaxies are slightly brighter than the cluster
ellipticals, but the difference implies a single-burst age difference
of at most 2.5 Gyr, and most likely smaller. Thus, it appears that
even if we include the disk-dominated galaxies in our analysis, the
results would not be significantly effected.

Fig.~9 also demonstrates that there is little apparent change
of FP slope with redshift. If galaxies evolved differently according to
their size, the residuals in Fig.~9 would show a trend with galaxy
size, assuming that we have applied the proper average evolutionary
correction.


\vskip 0.3cm \psfig{file=gebhardt.fig9.ps,width=8.8cm,angle=0}
\figcaption[fj2.ps]{The edge-on projection of the distant Fundamental
Plane adjusted for stellar luminosity evolution relation in
Fig.~5. Red points are galaxies with redshifts greater than 0.75; blue
points are below 0.75. The solid line is the best-fit calibration to
the local sample. The green open circles are galaxies that have bulge
fractions less than 0.4, which are considered late-type galaxies.
\label{fig9}}
\vskip 0.3cm


One goal for Fundamental Plane studies is to use luminosity evolution
to measure the stellar population formation history. From their study
of MS1054-03 at $z=0.83$ with a magnitude brightening of around 0.9
mag, van~Dokkum~\etal\ (1998) conclude that the star formation time
for this cluster is greater than $z=2.5$ assuming a single-burst
model. Our much larger brightening for field galaxies implies more
recent formation. To measure the change in galaxy brightness due to
stellar population aging, we use stellar population models of
Bruzual~\& Charlot (1998) and fit for the best-fit formation
redshift. The best-fit model for the field is shown as the red line in
Fig.~5. The formation redshifts for the field and cluster are 1.3 and
$>$3.0, respectively. However, as discussed next, by using only the
magnitude evolution we are ignoring an important constraint on
evolutionary models, namely the color. When color is introduced, the
simple single-burst stellar population model no longer fits the field
galaxy population.

\subsection{Stellar Population Models}


\begin{figure*}[t]
\vskip 0.3cm
\centerline{\psfig{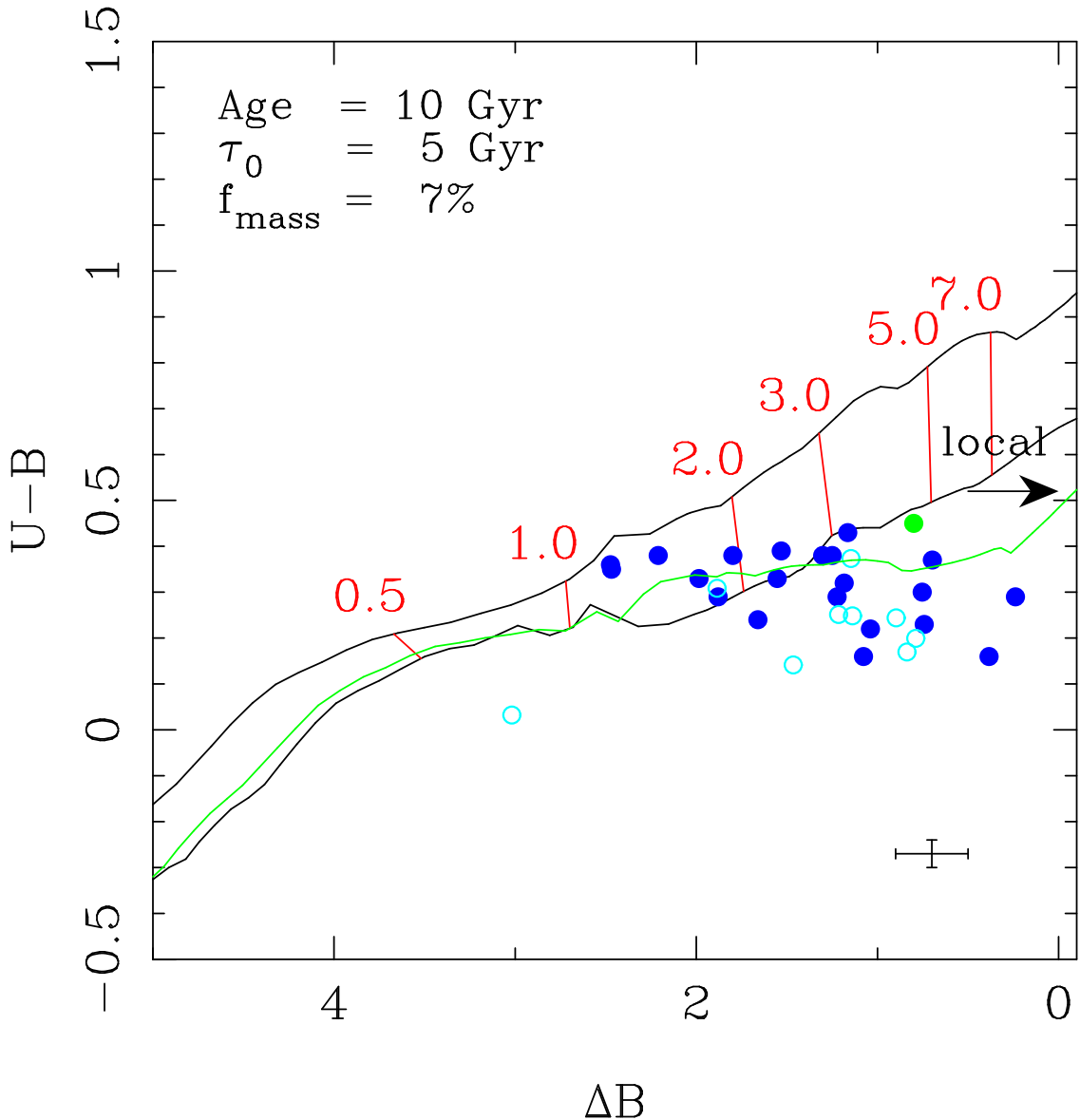}}
\figcaption[scombo.ps]{Rest-frame $U-B$ versus $B$ magnitude dimming,
the latter measured as residuals from the FP. The arrow labeled
``local'' is the mean integrated color of the local early-type sample
picked to match the distant field sample after fading (see Section 3
and Fig. 8). Blue points represent early-type field galaxy sample and
the one green point represents the cluster MS1054. The open symbols
are the field galaxies from van~Dokkum \& Ellis (2003). The error bars
in the bottom right represent typical 68\% uncertainties. The two
black lines refer to single-burst stellar population models with
metallicity solar and 2.5 times solar. The current age of both
galaxies is 10 Gyr in both cases, corresponding to a formation epoch
$z_f=1.9$, and the red labels correspond to the age of the galaxy at
that position. The green line results from including protracted star
formation assuming 2.5 solar metallicity and an exponentially falling
star-formation rate with an e-folding time of 5 Gyr. The total amount
of mass in this protracted component from the initial burst to the
present is 7\% of the total mass. Notice that in order to match the
very red colors despite large magnitude offsets, all models must start
with high metal content.
\label{fig10}}
\end{figure*}
\vskip 0.3cm


Colors provide another parameter which, when combined with luminosity
evolution, place tight limits on acceptable stellar population
models. Figure~10 plots the magnitude offset (from Fig.~5) versus
rest-frame $U-B$ color for early-types. As seen in both Fig.~10 and
Fig.~8, rest-frame colors evolve very little (less than 0.2 mag)
compared to local galaxies. Included in Fig. 10 are the points from
van Dokkum \& Ellis (2003), which show very similar $U-B$ colors but
less magnitude evolution. This amount of color change is not
consistent with any single metallicity and age. Shown in Fig.~10 are
two metallicities, solar and 2.5 times solar. Metallicities in between
these values span the curves shown in Fig.~10. The problem is to
reconcile the local red colors (around $U-B$ = 0.52) with the
similarly red colors (around 0.34) seen in distant galaxies that have
over two magnitudes of brightening.

Two plausible scenarios are population changes over time and changes
in dust contribution over time. Dust provides the mechanism to make
galaxies redder at high redshift if the dust is more significant in
the past. We know for local early-type galaxies that dust does have
appreciable effects (Falco~\etal\ 1999 and Goudfrooij~\& de~Jong
1995). However, the amount of dust required at high redshift to
explain the color change would cause us to notice its effect when
looking at galaxies at different inclinations. Koo~\etal\ (2003) find
consistent magnitudes and colors among their bulge samples between the
edge-on and face-on samples. Thus, we conclude that dust should not be
a major contribution to the reddening of these galaxies. Furthermore,
if dust were more prevalent at high redshifts, then the galaxies would
appear both redder and fainter. Dust would help to alleviate the
problem with having very red galaxies at high redshift, but it would
cause the galaxies to have an even more extreme magnitude brightening
which would push make the single-burst formation epoch of $z=1.5$ to
even later times.

An alternative approach is to invoke a changing stellar population.
The ways to do this include allowing for a contamination from stars
with different metallicities, different ages, or a combination of
both.  Either of these models can be made to work; however, in both,
the underlying requirement is to have galaxies that contain very red
colors at high-redshift. The only way to do this is to have an extreme
high-metallicity base component, or make the Universe old enough that
stars have the time to redden as they age. Since we have some basic
constraints on the age of the Universe from many cosmological studies,
and age constraints on local ellipticals (see Trager~\etal\ 2000 for
the most recent analysis), we are forced to use stars with
high-metallicity to explain the red colors. Thus, we always start with
a base population that has 2.5 times solar metallicity. Models by
either Bruzual \& Charlot (1998) or Worthey (1994) produce similar
results. One model includes accretion from a low metallicity
component. As seen in Fig.~10, low metallicity stars have a bluer
color, and thus their addition would serve to make the population
bluer. An acceptable accretion model (i.e., one that maintains
constant $U-B$) includes accreting stars with 1/50th solar metals that
are the same age as the base galaxy. The accretion is constant during
the time when the galaxy aged from 1-3 Gyrs, and the total mass
accreted is 10\%. Other models with different accretion histories will
also work, but the main requirement is that the accreted material have
very low metals if one uses the same age for both the base and the
accreted material. Unfortunately, it seems unreasonable to require the
accreted material to have such low metallicity when the base galaxy's
is so high. We therefore turn to star formation as a possible
mechanism.

As stars age they become redder. Thus, if we were to include stars
with younger ages, these blue stars would serve to make the whole
population bluer. We then simply adjust the amount of new stars to
make a null color change. The green line in Fig.~10 shows such a
model.  The base population consists of stars with 2.5 times solar
metals (the upper black line in the figure) that are 10 Gyr old
today. An exponentially declining star formation history provides the
necessary young stars to maintain constant color over time. The new
stars have solar metallicity, and the parameters are such that the
contribution today from the new stars is 7\% by mass and the e-folding
time for their formation is 5 Gyr. Keep in mind that most of the stars
are formed at early times; the age weighted by the $B$ light is 7.9
Gyr and weighted by the $U$ light is 5.7 Gyr, compared to the age of
the base population of 10 Gyr. These assumptions imply an intermediate
mass for the new population since we are making new stars from
solar-metallicity gas. Lower metals, being bluer, would require a
smaller amount of new stars; similarly, stars from higher metals would
require more new stars. Thus, if we use 2.5 times solar metals, we
find that 10\% of the total mass must be in new stars (with average
ages of 8.1 Gyr for the $B$ light and 6.2 Gyr for the $U$ light), and
using 1/5 solar metals requires only 4\% of the total mass. Any
metallicity can trace a nearly identical curve in Fig.~10 solely by
adjusting the fractional contribution of the new component. Thus,
given the large degeneracies and limited size of the present data, we
will not explore this further. However, for any metallicity the common
requirements are: 1) the base component must have high metals (2.5
times solar) otherwise we cannot achieve red enough colors at early
times; 2) the star formation rate must be peaked at early times,
otherwise the constancy of $U-B$ cannot be maintained; and 3) the
amount of late forming stars cannot exceed 10\%. The conclusion of 2)
above is largely independent of the stellar population model since it
results from a differential measurement. However, 1) and 3) rely on
good calibration of the stellar population models in an absolute sense
in the ultraviolet region, which is not very well tested. The fact
that the two sets of models---Bruzual \& Charlot and Worthey---give
similar results provides some comfort that the stellar population
models are accurate.

Local studies provide some insight into whether the above star
formation history is correct. Trager~\etal\ (2000a and 2000b)
investigate local field early-type galaxies to infer their stellar
population histories using structural and spectral parameters. They
find that local field ellipticals span a large range in
luminosity-weighted ages, with a slight tendency for lower-dispersion
galaxies to be younger. Furthermore, their best model suggests that
galaxies are made up of an older single stellar population with a
contribution from younger stars. This younger population appears to
come from pre-enriched gas rather than low-metallicity gas. While this
scenario is similar to our picture, the details differ, mainly in
regards to the metallicity of the base component. A second issue is
the $U$-band color, which has not yet been properly calibrated of
non-solar abundances ratios. The high metallicity---around 2.5
solar---required to achieve the very red colors at early times here
appears to be too high for the integrated metallicities measured for
local samples---generally from 1 to 1.5 solar. Resolution of this
difference may lie in a variety of places but possibly in the stellar
population models. At redshifts near 1, we are seeing galaxies as
young as or younger than 2 Gyr. These young ages pose a particularly
challenging task for stellar population models since the stellar
properties change rapidly during these times, and one has to be
careful about the input initial mass function, burst length, and many
other details which change depending on which model is
used. Broadhurst \& Bouwens (2000) demonstrate how changing the IMF
can cause very red colors for young galaxies--although at the expense
of a very truncated IMF. As the data increase and stellar population
models at young ages become more sophisticated, we will be in a good
position to re-examine these issues.

We can compare the above model of the star formation history for field
early-types to the model of van~Dokkum \& Franx (2001) for cluster
galaxies. They attempted to explain a similar problem that we have:
high redshift galaxies in clusters tend to have too large a magnitude
evolution for their constant, red colors. An additional problem is
that the spread of cluster colors is very small, too small to be
consistent with the cluster spread seen locally. Van~Dokkum \& Franx
invoke a model based on ``progenitor bias'' in which the galaxies in
their FP sample at high redshift do not represent the bulk of local
cluster galaxies today.  The galaxies that we observe in clusters
today have undergone significant evolution---from spirals to
ellipticals during merging---such that at redshifts near one, they are
not included in early type galaxies samples. Thus, the color spread
seen in high redshift ellipticals does not represent these today since
they represent only the oldest galaxies. Including the colors of the
progenitor spirals in the sample will increase the scatter and better
match the scatter seen in nearby cluster ellipticals. The consequence
is that when an elliptical is formed, it appears as a red galaxy since
the stars in the progenitor spirals are all formed at very high
redshifts ($z>3$). In their model, van~Dokkum \& Franx do not need to
invoke a changing stellar population to maintain a low scatter in the
galaxy colors. In order to explain their data, they suggest that 50\%
of the early-type galaxies in clusters are formed {\it after} a
redshift of one. If this model were also true for the field sample,
this large fraction of ellipticals would be seen in the evolution of
the luminosity function. Im~\etal\ (2002) measure essentially no
change in the number density of field ellipticals out to a redshift of
one. Their results exclude an increase in field ellipticals of 50\%
from $z=1$ to today by high significance (97\%). Thus, we cannot make
the cluster model of van~Dokkum \& Franx work for our sample.

We find that the field ellipticals have formation epochs at redshifts
less than 2, and likely around 1.5. The oldest cluster ellipticals
that are fully formed by $z=1$ have formation epochs $z>2.5$. The age
difference is around 2Gyr. We can see if this age difference is
possible from studies that compare local field and cluster
ellipticals. The most complete study comes from Bernardi~\etal\
(1998). Based on analysis of the Mg--$\sigma$ correlation, they find
little age difference between the field and cluster galaxies. They
suggest that field galaxies are on average $1.2\pm0.35$ Gyr younger
than cluster galaxies. However, as is shown by Trager et al. (2000a),
the Mg--$\sigma$ correlation is a composite of both age and
metallicity trends. Thus, it is difficult to draw definitive
conclusions so far about the local population. Another difficulty in
constraining age differences is that as a galaxy ages, the rate at
which it fades slows down. Thus, age differences between two old
galaxies (like those seen today) have a more subtle effect than in
young galaxies (such as at redshifts near one). Furthermore, the
Bernardi et al. cluster comparison sample represents the most extreme
forms of overdensities; i.e., they are some of the largest clusters
ever observed. Thus, if the galaxy formation epoch relates to the
local density, then the high redshift cluster/field comparison is
biased compared to the same at low redshift, where we have included a
much larger range of cluster sizes. Therefore, age differences between
the high redshift cluster/field sample are likely to be larger than
those seen at low redshift.


\vskip 0.3cm \psfig{file=gebhardt.fig11.ps,width=8.8cm,angle=0}
\figcaption[fig11.ps]{The relative contribution of giant stars in
best-fit template versus the magnitude offset from the Fundamental
Plane. The three bins which we used for the stellar template are A
dwarfs, F--G dwarfs, and G--K giants. There is a trend for galaxies
with small magnitude offsets to have larger contribution from more
evolved stars. Filled circles are early-types and open circles are
disk-dominated galaxies.
\label{fig11}}
\vskip 0.3cm


\subsection{Stellar Population Fraction}

We can also use the stellar templates that the spectral fitting
process chooses as probes of galaxy age. Many galaxies have spectra
covering regions from restframe 2500~\AA\ to 6000~\AA. These regions
include Balmer lines, which are good age indicators. Figure~2 and
Table~2 contain the relative fractions of stellar types. We divide the
template contributions into three bins by adding up the similar
stellar types in the two different sets of libraries. The three bins
include A stars, F--G dwarfs, and G--K giants. We plot the fraction of
giant stars against the magnitude offset from the Fundamental Plane in
Figure~11. There is a dramatic decrease of giant star contribution for
galaxies with larger magnitude offsets. This decrease corresponds to
an increase in the contribution of younger stars (the A-G dwarfs). We
find no significant trends between the A or F--G dwarfs. This trend is
obvious from inspection of the spectra. For those galaxies with large
contribution from young stars, we notice strong Balmer lines which are
absent in those galaxies dominated by giant star spectra. We could use
the relative contribution as an age indicator, similar to what we did
for the magnitude offset and color change in Figure~10; however, a
full treatment will have to wait until the stellar population models
have been scrutinized in the H+K region and the sample size increases,
since there is considerable noise in the estimate of the stellar
makeup. In any event, the correlation in Fig.~11 suggests that larger
magnitude offsets indicate younger stellar systems.

We have checked whether the correlation seen in Fig.~11 results from A
stars having broader lines in the H+K region. If, for example, the
fitting algorithm chooses A stars for a template as opposed to a giant
star, then the implied velocity dispersion would be smaller using the
A templates. This decrease in the width would cause a larger offset
from the FP, which could lead to the correlation in Fig.~11. However,
for those galaxies with low velocity dispersions, we have re-measured
dispersions using subsets of templates from A stars to giant stars. We
find little difference in the galaxy velocity dispersion. Furthermore,
the line ratios change dramatically from A dwarfs to K giants. The
fitting algorithm naturally takes this into account when it chooses
which stars provide the best fit.

Many A stars are rotationally broadened and we must consider whether
this impacts our measured dispersions. The study of Hubrig~\etal\
(2000) presents the distribution of $v~sin\,i$ for both young and old
stars. The peak of the distribution for both the old and young A-stars
is around 100~\kms, with a tail up to 250~\kms. If the galaxy light is
dominated by fast-rotating A stars and we do not include this in the
analysis, then we may overestimate the velocity
dispersion. Alternatively, we could underestimate the dispersion if
our template A star has systematically higher rotation. Neither of
these are important for our results. The $v~sin\,i$ for our template
stars are all less than 60~\kms. However, some of our galaxies are
dominated by A stars. We can estimate the effect on the dispersion
from the $v~sin\,i$ distribution. For a star rotating at 100~\kms, the
FWHM of a Balmer line will be around that value and most likely
smaller depending on the amount of limb darkening. Thus, the
dispersion due to rotation is around 40~\kms. Since all of the
measured galaxy dispersions are over 100~\kms, even if the galaxy is
completely dominated by A stars, the effect will be minimal. In order
for the A star rotation to have a significant impact on the galaxy,
{\it all} of the A stars in the galaxy would have to have rotations
that are at the extreme maximum of the known distribution. This is
unlikely and we conclude that rotation of the A stars is insignificant
for our purposes.

\subsection{Other Concerns}

One of the primary issues that arise when comparing redshifted samples
to local samples is whether we are looking at the same objects. The
local Fundamental Plane sample is comprised primarily of at least
spheroidal or early-type galaxies. We have been careful in our
analysis to highlight galaxies classified as early-type (58\% percent
of our sample); however there is uncertainty in this classification
system. To estimate the effect of this concern, we can use the
location in the Fundamental Plane of the disk-dominated galaxies.  The
15 disk-dominated galaxies provide very similar magnitude offsets and
colors as the 21 early-types. Thus, if we blindly compare these
distant, disk-dominated galaxies to the local early-type population,
we would arrive at similar conclusions. As seen in Fig.~5,
non-spheroidal galaxies have a slightly larger scatter at a given
redshift, but on average they follow the same magnitude
evolution. From Fig.~9, we see that the late-type galaxies (those that
have bulge fractions less than 0.4 and tend to show spiral structure)
tend to have very slightly less luminosity evolution, suggesting that
there is at most a small bias. Therefore, even if our classification
system includes some galaxies that would not be included locally, the
results would be unchanged.

A further complication in terms of the sample properties is that we
are using total galaxy properties instead of bulge properties, which
are likely to be more appropriate for FP studies. Koo~\etal\ (2003)
present a detailed analysis of bulge properties for a large sample of
objects that includes many of our galaxies. We refer to that paper
for a detailed discussion of bulge properties. Since we cannot
separate the velocity dispersion into contributions from the bulge and
disk light, we simply use total galaxy properties. However, even for
local Fundamental Plane studies, this decomposition is not used,
implying that our distant early-type sample should provide the best
comparison.

The large magnitude offsets from the Fundamental Plane are much larger
than any of our systematic effects, and are thus real. The relevant
parameters are $\sigma$, R$_e$, and surface brightness. Aperture
errors in surface brightness are at most 10\%, which implies 0.10
magnitudes. Systematic errors in R$_e$ are less than 10\% as
well. However, by plotting the combination of log~R$_e$ and log~SB
used here, the systematic errors are greatly reduced since the two
parameters are correlated. Kelson~\etal\ (2000) estimate that their
systematic errors on the FP offset from R$_e$ and surface brightness
is less than 3\% at $z=0.5$. Systematic errors in $\sigma$ are likely
to be less than 15\%, or 0.2 magnitudes in surface brightness.  Thus,
the largest uncertainties come from the measurement of $\sigma$, and
these are much smaller than the $\approx$2 magnitude differences seen
in Figure~5. 

Kochanek~\etal\ (2000) and Rusin et al. (2003) use strong
gravitational lensing from the CASTLES project to infer the velocity
dispersion of field lensing galaxies, and, hence, measure their
location in the Fundamental Plane. Van de Ven et al. (2003) follow up
on these measurements with a re-analysis. Rusin et al. have 28 lenses
out to redshifts of one, and find a significantly smaller magnitude
brightening than we do---inconsistent also with the results of
Im~\etal\ (2002) and Treu et al. (2002). Their offset is similar to
the average found by van~Dokkum~\etal\ (1998, 2001, 2003) for cluster
and field galaxies. Fortunately, we have one galaxy in common between
our sample and Rusin et al.'s: the quad lens 093\_2470 (HST14176+522),
at redshift 0.81. For this galaxy, we measure a velocity dispersion of
202~\kms, while their dispersion is 275~\kms, which is reflected in
the difference in magnitude brightening. Kochanek et al. (2000) have
four other galaxies (with redshift below 0.4) with both spectroscopic
and lens velocity dispersions; they find agreement between the two
estimates. Thus, it is not clear what the reasons are for the
differences. However, the dispersion difference for 093\_2470 is real,
and the next step is to include a larger sample of high-redshift
comparison objects. The re-analysis by van de Ven et al. (2003)
suggests even less of a difference between our results and the lensing
results. They estimate a mean formation redshift of
$1.8^{+1.4}_{-0.5}$, which is well within our range, although still at
earlier redshifts than what we measure. They further conclude that the
field lens galaxies are about 10-15\% younger than the cluster
galaxies. Rusin et al. (2003) note the difference between their
Fundamental Plane and that of Treu et al. (2002) as being due to their
mass-selected sample compared to the other luminosity-selected
samples, but this does not explain the difference seen in the direct
comparison above. However, this selection effect may be part of the
answer for why there are overall differences among various Fundamental
Plane studies.

\section{Conclusion}

We have conducted a study of the photometric and kinematic parameters
of distant early-type filed galaxies and compared to local
values. Assuming that radii and velocity dispersion do not evolve, we
find a progressive brightening in surface brightness back in time that
amounts to over two magnitudes at $z=1$. Integrated $U-B$ colors in
contrast evolve much less.  These findings broadly agree with
published results for distant {\it cluster} early-type galaxies except
that the surface-brightening effect in the field population is roughly
twice as large.

A consistent star-formation model that accounts for the evolution in
brightness but not in color involves an early burst of metal-rich
stars comprising $>$90\% of the mass, to which a younger population
comprising $<$10\% of the mass is added later with an e-folding time
of about five Gyr. This model is not unique but does imply that
early-type field galaxies consist mainly of old stars with a frosting
of younger stars formed later. Some evidence for a younger frosting
has indeed been found in spectral studies of local field early-type
galaxies.

Not answered bu this study is whether the younger stellar populations
forms smoothly, as assumed here, or in bursts. Also not answered is
the age of the oldest stars, i.e., the age of the initial period of
intense star formation. The mean light-weighted luminosities of our
field stellar populations at $z=1$ are sufficiently bright that it is
possible that the formation redshift could be as low as 1.4. If so,
red galaxies should begin to disappear from the field shortly before
$z=1$.  This trend should be detectable in deep K-band studies of
distant field galaxies that are currently in progress and will be
published soon. Thus, it is possible that the long-sought
transformation from blue proto-galaxies to mature red E/S0's may
shortly be detected, at least for field populations.

\acknowledgements

We are grateful for useful discussion with Ricardo Schiavon, Nicolas
Cardiel, Raja Guhathakurta, Dan Kelson, and Pieter van~Dokkum. An
anonymous referee offered excellent and thoughtful comments. KG is
supported by NASA through Hubble Fellowship grant HF-01090.01-97A
awarded by the Space Telescope Science Institute, which is operated by
the Association of the Universities for Research in Astronomy, Inc.,
for NASA under contract NAS 5-26555. We acknowledge support from NSF
grants AST-9529028 and AST-0071198.

\clearpage

\appendix

\section{K-corrections}


\vskip 0.3cm \psfig{file=gebhardt.figA12a.ps,width=8.8cm,angle=0}
\vskip -8.66cm \hskip 9.0cm
\psfig{file=gebhardt.figA12b.ps,width=8.8cm,angle=0}
\figcaption[gebhardt.figA12a.ps]{The data and parametric fits for
rest-frame $U-B$ colors (left three columns) and the K-corrections
(right three columns) for the local sample observed at different
redshifts. The solid points represent the data and the lines are the
parametric fits. The red line in the $z=0.8$ plot for $U-B$ is the
van~Dokkum et al. transformation. The open circle is discussed in the
text.
\label{fig12}}
\vskip 0.3cm


The plots on the left side of Figure~A12 are the computed rest-frame
$U-B$ color against observed $V-I$ color at the redshift listed in the
plot. The panels on the right-side are the $B$ magnitude K-corrections
as a function of observed $V-I$ color. To compute K-corrections and
rest-frame $U-B$ colors, we use the SED database provided by
Storchi-Bergmann, Calzetti, \& Kinney (see \\
http://www.stsci.edu/ftp/catalogs/nearby\_gal/sed.html), which covers
1200--7500~\AA\ using IUE and ground-based spectra for 99 galaxies, 43
of which have the full wavelength coverage. We place the specified
wavelength bandpass at the appropriate location in the galaxy
spectrum according to the redshift, and sum the flux. The filter
curves and the $V$ and $I$ zeropoints come from the {\it HST}
database. Since at redshifts around 0.8, observed $V-I$ is similar to
rest-frame $U-B$ (except for a zeropoint shift), we transform all
galaxies to rest-frame $U-B$ to minimize SED uncertainties. Thus, from
the 43 SEDs, we have rest-frame $U-B$ for each redshift and observed
$V-I$. Similarly, we also measure the K-corrections for rest-frame $B$
using observed $I$ and $V-I$. A polynomial fit for both estimates to
the SEDs produces the transformation for each of our galaxies. The
polynomial is as follows

\begin{eqnarray}
U-B & = & -0.8079-0.049752z-1.6232z^{2}+1.04067z^{3}+1.5294z^{4}
-0.41190z^{5}-0.56986z^{6} \nonumber \\
& & +(0.61591+1.07249z-2.2925z^{2}+1.3370z^{3})(V-I)+(0.280481-0.387205z+
0.043121z^{2})(V-I)^{2} 
\end{eqnarray}

\begin{eqnarray}
k_{B} & = & 0.0496+0.46057z+1.40430z^{2}-0.19436z^{3}-0.2232z^{4}-
0.36506z^{5}+0.17594z^{6} \nonumber \\
& & +(2.0532-2.8326z+1.05580z^{2}-0.67625z^{3})(V-I)+
(0.10826-0.68097z+0.61781z^{2})(V-I)^{2}
\end{eqnarray}

\noindent Figure~A12 shows the data and the polynomial fits. There is
one galaxy in the sample which is an obvious outlier---shown as an
open symbol, NGC~5128---which is excluded from the fit. Compared to
the results of Bruzual~\& Charlot (1998) models, there is acceptable
agreement, although we prefer to use the local spectra to provide the
intrinsic scatter. 

The comparison with the local color distribution of early-type
galaxies in Section 3.1 uses observed values for $U-B$, and we must
make sure that our system estimated here from SEDs is on the same
system as observed $U-B$'s. There are 15 galaxies in common between
the two samples. The average $U-B$ difference between the two is
$0.047\pm0.03$, with the measured SED $U-B$ being redder. The quoted
uncertainty is the 68\% confidence band on the mean offset. A possible
complication is that the aperture used for the SED $U-B$ values is not
the same that used for the local sample integrated colors. The
aperture used for the SED's is a $10\arcsec\times20\arcsec$ box for
both the IUE and ground-based spectra. Since this aperture does not
encompass the full galaxy, we expect the SED colors to be slightly
redder, as measured. However, the small measured difference between
the two colors is consistent with zero and is furthermore small for
the effects that we discuss.

\begin{deluxetable}{lccccccccccccccc}
\rotate
\tablecolumns{16}
\tablewidth{0pt}
\tablenum{1}
\tablecaption{Galaxy Sample}
\tablehead{
\colhead{(1)} &
\colhead{(2)} &
\colhead{(3)} &
\colhead{(4)} &
\colhead{(5)} &
\colhead{(6)} &
\colhead{(7)} &
\colhead{(8)} &
\colhead{(9)} &
\colhead{(10)}&
\colhead{(11)}&
\colhead{(12)}&
\colhead{(13)}&
\colhead{(14)}&
\colhead{(15)}&
\colhead{(16)}\\
\colhead{Galaxy}         & 
\colhead{z}              &
\colhead{Exptime}        &
\colhead{b$_{\rm f}$}    &
\colhead{m$_{I}$}        &
\colhead{M$_{B}$}        &
\colhead{$V-I$}          &
\colhead{$U-B$}          &
\colhead{r$_{1/2}$}      &
\colhead{r$_{1/2}$}      &
\colhead{SB$_e$}       &
\colhead{$\sigma$}       &
\colhead{$\Delta$SB}     &
\colhead{Type}           &
\colhead{RA}             &
\colhead{DEC}            \\
\colhead{Groth ID}       & 
\colhead{}               &
\colhead{sec}            &
\colhead{F814W}          &
\colhead{mag}            &
\colhead{mag}            &
\colhead{mag}            &
\colhead{mag}            &
\colhead{pixel}          &
\colhead{kpc}            &
\colhead{$B$ mags}       &
\colhead{\kms}           &
\colhead{$B$ mags}       &
\colhead{}               &
\colhead{J2000}          &
\colhead{J2000}          \\}
\startdata
052\_6543 &  0.7554 &  3000 &  0.44 &  20.31 &  --21.82 &  1.84 &  0.29 &  5.84 &  3.63 &  19.95 &  133 &  1.88 &  1 &  14:17:48.276 &  52:31:17.25 \\ 
	\nodata & \nodata & \nodata  & $_{-0.01}^{+0.02}$  & $_{-0.01}^{+0.01}$  & $_{-0.01}^{+0.01}$  & $_{-0.05}^{+0.04}$  & $_{-0.04}^{+0.03}$  & $_{-0.20}^{+0.16}$  & $_{-0.02}^{+0.01}$  & $_{-0.07}^{+0.07}$  & $_{-28}^{+22}$  & $_{-0.33}^{+0.26}$ & \nodata & \nodata & \nodata \\
062\_2060 &  0.9853 &  3000 &  0.44 &  21.16 &  --22.21 &  2.12 &  0.37 &  2.62 &  3.32 &  17.94 &  143 &  2.79 &  0 &  14:17:45.993 &  52:30:32.12 \\ 
	\nodata & \nodata & \nodata  & $_{-0.21}^{+0.06}$  & $_{-0.02}^{+0.02}$  & $_{-0.03}^{+0.03}$  & $_{-0.05}^{+0.05}$  & $_{-0.02}^{+0.02}$  & $_{-0.32}^{+0.29}$  & $_{-0.06}^{+0.05}$  & $_{-0.25}^{+0.19}$  & $_{-29}^{+23}$  & $_{-0.40}^{+0.31}$ & \nodata & \nodata & \nodata \\
062\_5761 &  0.6597 &  3000 &  0.49 &  19.96 &  --21.72 &  1.35 &  -0.02 &  6.47 &  3.65 &  20.12 &  147 &  1.61 &  0 &  14:17:42.150 &  52:30:25.24 \\ 
	\nodata & \nodata & \nodata  & $_{-0.02}^{+0.01}$  & $_{-0.01}^{+0.01}$  & $_{-0.01}^{+0.01}$  & $_{-0.02}^{+0.01}$  & $_{-0.02}^{+0.01}$  & $_{-0.36}^{+0.33}$  & $_{-0.02}^{+0.02}$  & $_{-0.08}^{+0.08}$  & $_{-19}^{+19}$  & $_{-0.21}^{+0.21}$ & \nodata & \nodata & \nodata \\
073\_2675 &  0.8086 &  7200 &  0.34 &  22.29 &  --20.14 &  2.02 &  0.37 &  1.46 &  3.04 &  18.70 &  127 &  1.36 &  0 &  14:17:49.318 &  52:29:07.54 \\ 
	\nodata & \nodata & \nodata  & $_{-0.02}^{+0.02}$  & $_{-0.01}^{+0.01}$  & $_{-0.01}^{+0.01}$  & $_{-0.03}^{+0.02}$  & $_{-0.02}^{+0.01}$  & $_{-0.09}^{+0.08}$  & $_{-0.03}^{+0.02}$  & $_{-0.12}^{+0.14}$  & $_{-29}^{+26}$  & $_{-0.37}^{+0.35}$ & \nodata & \nodata & \nodata \\
073\_3539 &  0.6440 &  3000 &  0.19 &  19.60 &  --21.95 &  1.53 &  0.15 &  7.61 &  3.72 &  20.22 &  221 &  1.02 &  0 &  14:17:45.384 &  52:29:08.27 \\ 
	\nodata & \nodata & \nodata  & $_{-0.00}^{+0.00}$  & $_{-0.00}^{+0.00}$  & $_{-0.00}^{+0.00}$  & $_{-0.01}^{+0.00}$  & $_{-0.01}^{+0.00}$  & $_{-0.04}^{+0.04}$  & $_{-0.00}^{+0.00}$  & $_{-0.01}^{+0.01}$  & $_{-23}^{+26}$  & $_{-0.16}^{+0.18}$ & \nodata & \nodata & \nodata \\
073\_7749 &  0.8728 &  6000 &  0.31 &  21.06 &  --21.70 &  1.96 &  0.30 &  2.35 &  3.26 &  18.20 &  161 &  2.14 &  0 &  14:17:45.473 &  52:29:50.81 \\ 
	\nodata & \nodata & \nodata  & $_{-0.01}^{+0.01}$  & $_{-0.01}^{+0.01}$  & $_{-0.01}^{+0.01}$  & $_{-0.01}^{+0.01}$  & $_{-0.01}^{+0.01}$  & $_{-0.09}^{+0.06}$  & $_{-0.02}^{+0.01}$  & $_{-0.06}^{+0.05}$  & $_{-31}^{+28}$  & $_{-0.30}^{+0.27}$ & \nodata & \nodata & \nodata \\
083\_5454 &  0.7553 &  3000 &  0.00 &  20.58 &  --21.57 &  1.37 &  -0.06 &  9.36 &  3.84 &  21.21 &  143 &  1.14 &  0 &  14:17:39.851 &  52:28:21.56 \\ 
	\nodata & \nodata & \nodata  & $_{-0.00}^{+0.00}$  & $_{-0.01}^{+0.02}$  & $_{-0.01}^{+0.02}$  & $_{-0.03}^{+0.03}$  & $_{-0.02}^{+0.02}$  & $_{-0.16}^{+0.21}$  & $_{-0.01}^{+0.01}$  & $_{-0.03}^{+0.04}$  & $_{-48}^{+44}$  & $_{-0.52}^{+0.48}$ & \nodata & \nodata & \nodata \\
083\_6766 &  0.4319 &  3600 &  0.53 &  21.00 &  --19.22 &  1.34 &  0.23 &  1.87 &  3.02 &  19.47 &  112 &  0.74 &  1 &  14:17:40.909 &  52:28:35.90 \\ 
	\nodata & \nodata & \nodata  & $_{-0.06}^{+0.07}$  & $_{-0.01}^{+0.01}$  & $_{-0.02}^{+0.02}$  & $_{-0.03}^{+0.03}$  & $_{-0.03}^{+0.03}$  & $_{-0.46}^{+0.30}$  & $_{-0.12}^{+0.06}$  & $_{-0.46}^{+0.22}$  & $_{-18}^{+18}$  & $_{-0.52}^{+0.33}$ & \nodata & \nodata & \nodata \\
084\_1138 &  0.8122 &  5400 &  0.45 &  20.67 &  --21.76 &  1.55 &  0.05 &  6.03 &  3.66 &  20.14 &  124 &  1.89 &  0 &  14:17:37.255 &  52:26:49.66 \\ 
	\nodata & \nodata & \nodata  & $_{-0.02}^{+0.02}$  & $_{-0.02}^{+0.02}$  & $_{-0.03}^{+0.02}$  & $_{-0.05}^{+0.05}$  & $_{-0.03}^{+0.03}$  & $_{-0.56}^{+0.39}$  & $_{-0.04}^{+0.03}$  & $_{-0.16}^{+0.14}$  & $_{-31}^{+29}$  & $_{-0.42}^{+0.39}$ & \nodata & \nodata & \nodata \\
084\_4521 &  0.7540 &  3000 &  0.03 &  20.50 &  --21.65 &  1.43 &  -0.01 &  5.49 &  3.60 &  19.97 &  182 &  1.24 &  0 &  14:17:40.551 &  52:27:13.59 \\ 
	\nodata & \nodata & \nodata  & $_{-0.00}^{+0.00}$  & $_{-0.01}^{+0.01}$  & $_{-0.01}^{+0.01}$  & $_{-0.02}^{+0.02}$  & $_{-0.01}^{+0.01}$  & $_{-0.10}^{+0.24}$  & $_{-0.01}^{+0.02}$  & $_{-0.04}^{+0.08}$  & $_{-35}^{+16}$  & $_{-0.30}^{+0.16}$ & \nodata & \nodata & \nodata \\
092\_2023 &  0.9874 &  13800 &  0.84 &  21.73 &  --21.63 &  2.09 &  0.36 &  0.80 &  2.80 &  15.97 &  213 &  2.47 &  1 &  14:17:27.256 &  52:26:27.51 \\ 
	\nodata & \nodata & \nodata  & $_{-0.06}^{+0.08}$  & $_{-0.02}^{+0.02}$  & $_{-0.02}^{+0.02}$  & $_{-0.04}^{+0.02}$  & $_{-0.02}^{+0.01}$  & $_{-0.18}^{+0.18}$  & $_{-0.11}^{+0.09}$  & $_{-0.40}^{+0.27}$  & $_{-39}^{+44}$  & $_{-0.49}^{+0.42}$ & \nodata & \nodata & \nodata \\
092\_4946 &  0.5335 &  6000 &  0.01 &  20.82 &  --20.39 &  0.88 &  -0.33 &  3.84 &  3.38 &  20.05 &  158 &  0.71 &  0 &  14:17:23.584 &  52:26:43.83 \\ 
	\nodata & \nodata & \nodata  & $_{-0.01}^{+0.01}$  & $_{-0.01}^{+0.01}$  & $_{-0.01}^{+0.01}$  & $_{-0.02}^{+0.01}$  & $_{-0.02}^{+0.01}$  & $_{-0.08}^{+0.08}$  & $_{-0.01}^{+0.01}$  & $_{-0.04}^{+0.04}$  & $_{-14}^{+4}$  & $_{-0.14}^{+0.06}$ & \nodata & \nodata & \nodata \\
093\_2470 &  0.8110 &  18600 &  0.52 &  19.69 &  --22.74 &  1.83 &  0.24 &  9.39 &  3.85 &  20.14 &  202 &  1.66 &  1 &  14:17:35.745 &  52:26:45.66 \\ 
	\nodata & \nodata & \nodata  & $_{-0.02}^{+0.04}$  & $_{-0.02}^{+0.02}$  & $_{-0.02}^{+0.02}$  & $_{-0.03}^{+0.04}$  & $_{-0.02}^{+0.03}$  & $_{-0.53}^{+0.75}$  & $_{-0.03}^{+0.03}$  & $_{-0.11}^{+0.16}$  & $_{-10}^{+8}$  & $_{-0.13}^{+0.17}$ & \nodata & \nodata & \nodata \\
094\_2660 &  0.9033 &  5400 &  0.42 &  20.69 &  --22.25 &  2.12 &  0.38 &  7.77 &  3.78 &  20.25 &  154 &  1.80 &  1 &  14:17:32.870 &  52:25:21.72 \\ 
	\nodata & \nodata & \nodata  & $_{-0.02}^{+0.02}$  & $_{-0.02}^{+0.03}$  & $_{-0.03}^{+0.03}$  & $_{-0.08}^{+0.09}$  & $_{-0.04}^{+0.05}$  & $_{-0.58}^{+0.48}$  & $_{-0.03}^{+0.03}$  & $_{-0.14}^{+0.12}$  & $_{-39}^{+35}$  & $_{-0.41}^{+0.37}$ & \nodata & \nodata & \nodata \\
102\_3649 &  0.5336 &  3000 &  0.76 &  21.17 &  --19.64 &  1.62 &  0.38 &  1.43 &  2.95 &  18.71 &  115 &  1.25 &  1 &  14:17:18.412 &  52:25:39.05 \\ 
	\nodata & \nodata & \nodata  & $_{-0.02}^{+0.01}$  & $_{-0.02}^{+0.01}$  & $_{-0.02}^{+0.02}$  & $_{-0.03}^{+0.03}$  & $_{-0.03}^{+0.03}$  & $_{-0.11}^{+0.10}$  & $_{-0.03}^{+0.03}$  & $_{-0.12}^{+0.11}$  & $_{-11}^{+17}$  & $_{-0.19}^{+0.25}$ & \nodata & \nodata & \nodata \\
103\_1115 &  0.4635 &  6000 &  0.76 &  19.91 &  --20.49 &  1.44 &  0.29 &  4.22 &  3.39 &  20.06 &  118 &  1.22 &  1 &  14:17:23.695 &  52:25:11.28 \\ 
	\nodata & \nodata & \nodata  & $_{-0.02}^{+0.02}$  & $_{-0.02}^{+0.01}$  & $_{-0.02}^{+0.01}$  & $_{-0.03}^{+0.02}$  & $_{-0.03}^{+0.02}$  & $_{-0.22}^{+0.32}$  & $_{-0.02}^{+0.03}$  & $_{-0.11}^{+0.14}$  & $_{-7}^{+9}$  & $_{-0.14}^{+0.18}$ & \nodata & \nodata & \nodata \\
103\_2074 &  1.0234 &  7200 &  0.82 &  21.79 &  --21.83 &  2.42 &  0.51 &  4.15 &  3.52 &  19.30 &  134 &  2.18 &  0 &  14:17:29.697 &  52:25:32.53 \\ 
	\nodata & \nodata & \nodata  & $_{-0.18}^{+0.08}$  & $_{-0.04}^{+0.04}$  & $_{-0.05}^{+0.05}$  & $_{-0.07}^{+0.10}$  & $_{-0.03}^{+0.04}$  & $_{-1.06}^{+1.17}$  & $_{-0.13}^{+0.11}$  & $_{-0.43}^{+0.39}$  & $_{-41}^{+43}$  & $_{-0.64}^{+0.63}$ & \nodata & \nodata & \nodata \\
103\_4766 &  0.8118 &  3000 &  0.45 &  21.10 &  --21.35 &  2.04 &  0.38 &  1.19 &  2.95 &  17.05 &  173 &  2.21 &  1 &  14:17:28.264 &  52:25:57.20 \\ 
	\nodata & \nodata & \nodata  & $_{-0.02}^{+0.02}$  & $_{-0.01}^{+0.01}$  & $_{-0.01}^{+0.01}$  & $_{-0.02}^{+0.03}$  & $_{-0.01}^{+0.02}$  & $_{-0.18}^{+0.17}$  & $_{-0.07}^{+0.06}$  & $_{-0.23}^{+0.18}$  & $_{-30}^{+30}$  & $_{-0.35}^{+0.32}$ & \nodata & \nodata & \nodata \\
\tablebreak
103\_7221 &  0.9013 &  10800 &  0.54 &  20.79 &  --22.14 &  2.06 &  0.35 &  6.43 &  3.70 &  19.94 &  107 &  2.47 &  1 &  14:17:22.868 &  52:26:10.99 \\ 
	\nodata & \nodata & \nodata  & $_{-0.05}^{+0.06}$  & $_{-0.03}^{+0.03}$  & $_{-0.04}^{+0.04}$  & $_{-0.09}^{+0.11}$  & $_{-0.05}^{+0.06}$  & $_{-0.47}^{+0.47}$  & $_{-0.03}^{+0.03}$  & $_{-0.14}^{+0.17}$  & $_{-55}^{+57}$  & $_{-0.81}^{+0.84}$ & \nodata & \nodata & \nodata \\
113\_3311 &  0.8117 &  16500 &  0.77 &  20.50 &  --21.95 &  1.97 &  0.33 &  3.99 &  3.48 &  19.07 &  158 &  1.99 &  1 &  14:17:16.219 &  52:24:21.27 \\ 
	\nodata & \nodata & \nodata  & $_{-0.01}^{+0.02}$  & $_{-0.02}^{+0.02}$  & $_{-0.02}^{+0.02}$  & $_{-0.04}^{+0.05}$  & $_{-0.03}^{+0.03}$  & $_{-0.27}^{+0.41}$  & $_{-0.03}^{+0.04}$  & $_{-0.11}^{+0.17}$  & $_{-10}^{+11}$  & $_{-0.15}^{+0.20}$ & \nodata & \nodata & \nodata \\
113\_4933 &  0.8112 &  6000 &  0.02 &  21.42 &  --21.00 &  1.31 &  -0.11 &  2.66 &  3.30 &  19.11 &  166 &  1.31 &  0 &  14:17:18.134 &  52:24:41.82 \\ 
	\nodata & \nodata & \nodata  & $_{-0.02}^{+0.05}$  & $_{-0.02}^{+0.02}$  & $_{-0.02}^{+0.02}$  & $_{-0.02}^{+0.03}$  & $_{-0.01}^{+0.02}$  & $_{-0.09}^{+0.05}$  & $_{-0.01}^{+0.01}$  & $_{-0.07}^{+0.05}$  & $_{-23}^{+4}$  & $_{-0.23}^{+0.06}$ & \nodata & \nodata & \nodata \\
172\_7753 &  0.6480 &  3600 &  0.86 &  20.38 &  --21.12 &  1.81 &  0.39 &  1.73 &  3.08 &  17.89 &  198 &  1.53 &  1 &  14:16:28.442 &  52:17:23.93 \\ 
	\nodata & \nodata & \nodata  & $_{-0.01}^{+0.01}$  & $_{-0.01}^{+0.02}$  & $_{-0.01}^{+0.01}$  & $_{-0.04}^{+0.02}$  & $_{-0.04}^{+0.02}$  & $_{-0.12}^{+0.16}$  & $_{-0.03}^{+0.04}$  & $_{-0.11}^{+0.14}$  & $_{-35}^{+38}$  & $_{-0.29}^{+0.33}$ & \nodata & \nodata & \nodata \\
192\_3330 &  0.5759 &  7200 &  0.41 &  20.61 &  --20.48 &  1.64 &  0.33 &  1.42 &  2.97 &  17.93 &  156 &  1.55 &  1 &  14:16:20.754 &  52:14:52.35 \\ 
	\nodata & \nodata & \nodata  & $_{-0.03}^{+0.02}$  & $_{-0.02}^{+0.01}$  & $_{-0.01}^{+0.01}$  & $_{-0.02}^{+0.03}$  & $_{-0.02}^{+0.03}$  & $_{-0.21}^{+0.16}$  & $_{-0.07}^{+0.05}$  & $_{-0.22}^{+0.18}$  & $_{-17}^{+20}$  & $_{-0.28}^{+0.27}$ & \nodata & \nodata & \nodata \\
193\_1227 &  0.7992 &  3600 &  0.73 &  21.18 &  --21.20 &  2.01 &  0.37 &  3.20 &  3.38 &  19.35 &  239 &  0.70 &  1 &  14:16:26.467 &  52:14:45.93 \\ 
	\nodata & \nodata & \nodata  & $_{-0.05}^{+0.05}$  & $_{-0.03}^{+0.04}$  & $_{-0.03}^{+0.04}$  & $_{-0.09}^{+0.08}$  & $_{-0.06}^{+0.05}$  & $_{-0.40}^{+0.43}$  & $_{-0.06}^{+0.05}$  & $_{-0.23}^{+0.27}$  & $_{-32}^{+37}$  & $_{-0.31}^{+0.36}$ & \nodata & \nodata & \nodata \\
203\_1613 &  0.6404 &  3600 &  0.38 &  20.39 &  --21.10 &  1.65 &  0.26 &  4.91 &  3.53 &  20.12 &  129 &  1.44 &  0 &  14:16:18.384 &  52:13:35.18 \\ 
	\nodata & \nodata & \nodata  & $_{-0.02}^{+0.03}$  & $_{-0.02}^{+0.02}$  & $_{-0.01}^{+0.02}$  & $_{-0.04}^{+0.05}$  & $_{-0.04}^{+0.04}$  & $_{-0.21}^{+0.28}$  & $_{-0.02}^{+0.02}$  & $_{-0.09}^{+0.11}$  & $_{-21}^{+26}$  & $_{-0.27}^{+0.33}$ & \nodata & \nodata & \nodata \\
203\_2634 &  0.3660 &  3600 &  0.83 &  17.83 &  --21.85 &  1.32 &  0.29 &  9.57 &  3.68 &  20.23 &  328 &  0.24 &  1 &  14:16:20.355 &  52:13:49.30 \\ 
	\nodata & \nodata & \nodata  & $_{-0.01}^{+0.00}$  & $_{-0.02}^{+0.02}$  & $_{-0.02}^{+0.02}$  & $_{-0.02}^{+0.02}$  & $_{-0.02}^{+0.02}$  & $_{-0.12}^{+0.12}$  & $_{-0.01}^{+0.01}$  & $_{-0.06}^{+0.07}$  & $_{-24}^{+19}$  & $_{-0.13}^{+0.11}$ & \nodata & \nodata & \nodata \\
203\_4339 &  0.8126 &  3600 &  0.27 &  20.38 &  --22.06 &  1.95 &  0.32 &  5.01 &  3.57 &  19.48 &  197 &  1.51 &  0 &  14:16:20.515 &  52:14:07.37 \\ 
	\nodata & \nodata & \nodata  & $_{-0.01}^{+0.02}$  & $_{-0.01}^{+0.02}$  & $_{-0.02}^{+0.02}$  & $_{-0.05}^{+0.04}$  & $_{-0.03}^{+0.03}$  & $_{-0.21}^{+0.18}$  & $_{-0.02}^{+0.02}$  & $_{-0.09}^{+0.08}$  & $_{-23}^{+20}$  & $_{-0.20}^{+0.17}$ & \nodata & \nodata & \nodata \\
212\_4836 &  0.3522 &  3600 &  0.51 &  19.35 &  --20.29 &  1.24 &  0.22 &  2.72 &  3.13 &  18.97 &  154 &  1.04 &  1 &  14:16:06.059 &  52:12:33.78 \\ 
	\nodata & \nodata & \nodata  & $_{-0.01}^{+0.01}$  & $_{-0.02}^{+0.02}$  & $_{-0.02}^{+0.02}$  & $_{-0.02}^{+0.02}$  & $_{-0.02}^{+0.02}$  & $_{-0.11}^{+0.08}$  & $_{-0.02}^{+0.01}$  & $_{-0.06}^{+0.05}$  & $_{-6}^{+6}$  & $_{-0.09}^{+0.08}$ & \nodata & \nodata & \nodata \\
262\_7430 &  0.6064 &  3600 &  0.83 &  20.98 &  --20.25 &  1.79 &  0.43 &  1.53 &  3.01 &  18.38 &  163 &  1.16 &  1 &  14:15:31.190 &  52:06:32.89 \\ 
	\nodata & \nodata & \nodata  & $_{-0.03}^{+0.03}$  & $_{-0.02}^{+0.04}$  & $_{-0.02}^{+0.03}$  & $_{-0.05}^{+0.04}$  & $_{-0.05}^{+0.04}$  & $_{-0.12}^{+0.10}$  & $_{-0.04}^{+0.03}$  & $_{-0.13}^{+0.12}$  & $_{-35}^{+29}$  & $_{-0.36}^{+0.30}$ & \nodata & \nodata & \nodata \\
274\_3875 &  0.2826 &  3600 &  0.84 &  19.49 &  --19.59 &  1.12 &  0.16 &  2.39 &  3.01 &  19.07 &  171 &  0.38 &  1 &  14:15:37.949 &  52:04:08.22 \\ 
	\nodata & \nodata & \nodata  & $_{-0.01}^{+0.01}$  & $_{-0.02}^{+0.02}$  & $_{-0.02}^{+0.02}$  & $_{-0.02}^{+0.02}$  & $_{-0.02}^{+0.02}$  & $_{-0.05}^{+0.06}$  & $_{-0.01}^{+0.01}$  & $_{-0.04}^{+0.05}$  & $_{-11}^{+8}$  & $_{-0.11}^{+0.09}$ & \nodata & \nodata & \nodata \\
274\_5920 &  0.8110 &  3600 &  0.48 &  19.63 &  --22.82 &  2.03 &  0.38 &  10.41 &  3.89 &  20.31 &  245 &  1.30 &  1 &  14:15:38.889 &  52:05:06.56 \\ 
	\nodata & \nodata & \nodata  & $_{-0.01}^{+0.02}$  & $_{-0.03}^{+0.02}$  & $_{-0.03}^{+0.02}$  & $_{-0.06}^{+0.06}$  & $_{-0.04}^{+0.04}$  & $_{-0.30}^{+0.31}$  & $_{-0.01}^{+0.01}$  & $_{-0.06}^{+0.07}$  & $_{-35}^{+41}$  & $_{-0.23}^{+0.27}$ & \nodata & \nodata & \nodata \\
282\_2474 &  0.6315 &  3600 &  0.30 &  21.09 &  --20.55 &  1.02 &  -0.27 &  3.00 &  3.31 &  19.54 &  169 &  0.88 &  0 &  14:15:22.654 &  52:05:04.57 \\ 
	\nodata & \nodata & \nodata  & $_{-0.05}^{+0.05}$  & $_{-0.03}^{+0.04}$  & $_{-0.02}^{+0.03}$  & $_{-0.05}^{+0.05}$  & $_{-0.04}^{+0.04}$  & $_{-0.19}^{+0.18}$  & $_{-0.03}^{+0.03}$  & $_{-0.13}^{+0.12}$  & $_{-29}^{+4}$  & $_{-0.29}^{+0.12}$ & \nodata & \nodata & \nodata \\
283\_2254 &  0.6504 &  3600 &  0.82 &  20.14 &  --21.39 &  1.74 &  0.32 &  1.89 &  3.11 &  17.79 &  277 &  1.18 &  1 &  14:15:31.112 &  52:04:28.73 \\ 
	\nodata & \nodata & \nodata  & $_{-0.01}^{+0.02}$  & $_{-0.01}^{+0.02}$  & $_{-0.01}^{+0.01}$  & $_{-0.02}^{+0.03}$  & $_{-0.02}^{+0.03}$  & $_{-0.09}^{+0.07}$  & $_{-0.02}^{+0.02}$  & $_{-0.07}^{+0.05}$  & $_{-24}^{+23}$  & $_{-0.15}^{+0.14}$ & \nodata & \nodata & \nodata \\
\tablebreak
\tablebreak
\tablebreak
\tablebreak
\tablebreak
\tablebreak
\tablebreak
\tablebreak
283\_6152 &  0.8086 &  2080 &  0.39 &  20.79 &  --21.64 &  2.04 &  0.38 &  1.32 &  2.99 &  16.96 &  335 &  1.32 &  0 &  14:15:29.997 &  52:05:06.11 \\ 
	\nodata & \nodata & \nodata  & $_{-0.02}^{+0.04}$  & $_{-0.01}^{+0.02}$  & $_{-0.02}^{+0.02}$  & $_{-0.03}^{+0.02}$  & $_{-0.02}^{+0.01}$  & $_{-0.28}^{+0.21}$  & $_{-0.10}^{+0.06}$  & $_{-0.32}^{+0.21}$  & $_{-48}^{+53}$  & $_{-0.39}^{+0.32}$ & \nodata & \nodata & \nodata \\
284\_3854 &  0.4495 &  7200 &  0.79 &  19.56 &  --20.82 &  1.30 &  0.16 &  2.78 &  3.20 &  18.79 &  192 &  1.08 &  1 &  14:15:31.060 &  52:03:19.19 \\ 
	\nodata & \nodata & \nodata  & $_{-0.01}^{+0.01}$  & $_{-0.01}^{+0.02}$  & $_{-0.01}^{+0.02}$  & $_{-0.02}^{+0.03}$  & $_{-0.02}^{+0.03}$  & $_{-0.10}^{+0.09}$  & $_{-0.02}^{+0.01}$  & $_{-0.06}^{+0.05}$  & $_{-8}^{+7}$  & $_{-0.09}^{+0.07}$ & \nodata & \nodata & \nodata \\
313\_1250 &  0.5148 &  3600 &  0.87 &  20.14 &  --20.58 &  1.52 &  0.30 &  3.49 &  3.33 &  19.68 &  175 &  0.75 &  1 &  14:15:11.650 &  52:00:46.36 \\ 
	\nodata & \nodata & \nodata  & $_{-0.02}^{+0.03}$  & $_{-0.02}^{+0.02}$  & $_{-0.02}^{+0.02}$  & $_{-0.03}^{+0.04}$  & $_{-0.03}^{+0.04}$  & $_{-0.19}^{+0.28}$  & $_{-0.02}^{+0.03}$  & $_{-0.14}^{+0.12}$  & $_{-26}^{+25}$  & $_{-0.27}^{+0.25}$ & \nodata & \nodata & \nodata \\
\enddata 

\tablecomments{All rest frame quantities use H$_0=70,
\Omega_\Lambda=0.7, \Omega_{\rm m}=0.3$. The columns are galaxy Groth
ID (1), redshift (2), total exposure time for the spectroscopy (3),
bulge fraction in the filter F814W (4), observed total apparent
$I$-band Vega magnitude (5), total absolute $B$-band magnitude using
$k$-correction outlined in the Appendix (6), observed total $V-I$
color (7), calculated total $U-B$ color based on fit outlined in the
Appendix (8), observed half light radius in WFPC pixels, 0.0996\arcsec
(9), half-light radius in kpc (10), observed average $B$-band surface
brightness within the half-light radius (11), observed velocity
dispersion (not including aperture corrections) (12), calculated
surface brightness $B$-band magnitude offset from the Fundamental
Plane (13), type refers to whether a galaxy is considered a
disk-dominated (0) or an early-type (1) galaxy as classified in Im et
al. (2002) (14), and RA (15) and DEC (16). 68\% confidence bands are
provided under each of the measured quantities.}
\end{deluxetable}

\vfill\eject

\begin{deluxetable}{lccccccc}
\tablecolumns{8}
\tablewidth{0pt}
\tablenum{2}
\tablecaption{Spectral Data}
\tablehead{
\colhead{(1)} &
\colhead{(2)} &
\colhead{(3)} &
\colhead{(4)} &
\colhead{(5)} &
\colhead{(6)} &
\colhead{(7)} &
\colhead{(8)} \\
\colhead{Galaxy}         & 
\colhead{Region}         &
\colhead{S/N}            &
\colhead{\AA/pixel}      &
\colhead{Template}       &
\colhead{f$_{\rm AV}$}    &
\colhead{f$_{\rm F-GV}$}  &
\colhead{f$_{\rm G-KIII}$}\\
\colhead{Groth ID}       & 
\colhead{\AA}            &
\colhead{per pixel}      &
\colhead{\AA}            &
\colhead{}               &
\colhead{}               &
\colhead{}               &
\colhead{}               \\}
\startdata
052\_6543 &  $3850-4100$ &  12 & 1.20 &  KPNO &  0.22 &  0.78 &  0.00 \\
062\_2060 &  $3850-4200$ &  15 & 1.20 &  KG   &  0.28 &  0.72 &  0.00 \\
062\_5761 &  $3850-4100$ &  19 & 0.84 &  KG   &  0.59 &  0.41 &  0.00 \\
073\_2675 &  $3850-4050$ &   6 & 1.20 &  KPNO &  0.00 &  0.78 &  0.22 \\
073\_3539 &  $3850-4150$ &  18 & 1.20 &  KPNO &  0.20 &  0.80 &  0.00 \\
073\_7749 &  $3850-4100$ &  14 & 1.20 &  KPNO &  0.62 &  0.38 &  0.00 \\
083\_5454 &  $3900-4150$ &   9 & 1.20 &  KPNO &  0.52 &  0.00 &  0.48 \\
083\_6766 &  $3850-4505$ &   6 & 0.84 &  KG   &  0.00 &  0.00 &  1.00 \\
084\_1138 &  $3870-4200$ &  12 & 1.20 &  KPNO &  0.06 &  0.94 &  0.00 \\
084\_4521 &  $3850-4100$ &  14 & 1.20 &  KPNO &  0.48 &  0.00 &  0.52 \\
092\_2023 &  $3890-4155$ &  11 & 1.20 &  KG   &  0.10 &  0.90 &  0.00 \\
092\_4946 &  $3850-4100$ &  19 & 1.20 &  KPNO &  0.70 &  0.00 &  0.30 \\
093\_2470 &  $3850-4500$ &  28 & 1.20 &  KPNO &  0.02 &  0.79 &  0.19 \\
094\_2660 &  $3870-4100$ &  14 & 1.20 &  KPNO &  0.09 &  0.91 &  0.00 \\
102\_3649 &  $3850-4500$ &  11 & 1.20 &  KPNO &  0.16 &  0.59 &  0.25 \\
103\_1115 &  $3850-4500$ &  15 & 0.84 &  KPNO &  0.05 &  0.75 &  0.20 \\
103\_2074 &  $3850-4100$ &   6 & 1.20 &  KG   &  0.43 &  0.50 &  0.07 \\
103\_4766 &  $3850-4050$ &  11 & 1.20 &  KPNO &  0.03 &  0.97 &  0.00 \\
103\_7221 &  $3850-4050$ &   9 & 1.20 &  KG   &  0.18 &  0.82 &  0.00 \\
113\_3311 &  $3850-4500$ &  22 & 1.20 &  KPNO &  0.34 &  0.40 &  0.26 \\
113\_4933 &  $3850-4050$ &  17 & 1.28 &  KPNO &  0.75 &  0.00 &  0.25 \\
172\_7753 &  $3850-4075$ &  11 & 1.28 &  KG   &  0.12 &  0.59 &  0.29 \\
192\_3330 &  $3850-4100$ &  10 & 1.28 &  KG   &  0.00 &  0.79 &  0.21 \\
193\_1227 &  $3850-4050$ &   7 & 1.28 &  KPNO &  0.19 &  0.16 &  0.65 \\
203\_1613 &  $3850-4200$ &  13 & 1.28 &  KPNO &  0.11 &  0.53 &  0.36 \\
203\_2634 &  $3890-4400$ &  24 & 1.28 &  KPNO &  0.00 &  0.00 &  1.00 \\
203\_4339 &  $3850-4100$ &  18 & 1.28 &  KPNO &  0.18 &  0.82 &  0.00 \\
212\_4836 &  $3850-4500$ &  22 & 1.28 &  KPNO &  0.07 &  0.53 &  0.40 \\
262\_7430 &  $3850-4115$ &  10 & 1.28 &  KG   &  0.10 &  0.61 &  0.29 \\
274\_3875 &  $4000-4500$ &  23 & 1.28 &  KPNO &  0.11 &  0.33 &  0.56 \\
274\_5920 &  $3850-4100$ &   9 & 1.28 &  KPNO &  0.14 &  0.31 &  0.55 \\
282\_2474 &  $3850-4100$ &  13 & 1.28 &  KPNO &  0.74 &  0.00 &  0.26 \\
283\_2254 &  $3850-4450$ &  14 & 1.28 &  KPNO &  0.06 &  0.47 &  0.47 \\
283\_6152 &  $3850-4105$ &   8 & 1.28 &  KPNO &  0.49 &  0.00 &  0.51 \\
284\_3854 &  $3850-4500$ &  18 & 1.28 &  KG   &  0.00 &  0.77 &  0.23 \\
313\_1250 &  $3880-4100$ &  10 & 1.28 &  KG   &  0.10 &  0.45 &  0.45 \\
\enddata

\tablecomments{The wavelength region (column 2) corresponds to the
restframe spectral region that was used in the kinematic
extractions. It is different for each galaxy since the galaxy
redshifts determine which regions to exclude. The S/N is calculated
per pixel and the pixel size is given in column 4. Column 5 refers to
whether we used either the KPNO or the Kobulnicky \& Gebhardt (KG)
template library. Columns 6, 7, and 8 refer to the fractional
contribution of A dwarfs, F--G dwarfs, and G--K giants.}

\end{deluxetable}

\end{document}